\documentclass[9pt,sigconf]{acmart}
    
\usepackage{booktabs} 
\usepackage{CJK}
\usepackage{algorithmicx}
\usepackage[ruled,vlined,linesnumbered]{algorithm2e}
\usepackage{hhline}
\usepackage{amsmath,mathtools}
\usepackage{amsfonts}
\usepackage{mathrsfs}
\usepackage{graphicx}
\usepackage{subcaption}
\usepackage{enumitem,color}

\copyrightyear{2021}
\acmYear{2021}
\setcopyright{acmcopyright}\acmConference[CoNEXT '21]{The 17th International Conference on emerging Networking EXperiments and Technologies}{December 7--10, 2021}{Virtual Event, Germany}
\acmBooktitle{The 17th International Conference on emerging Networking EXperiments and Technologies (CoNEXT '21), December 7--10, 2021, Virtual Event, Germany}
\acmPrice{15.00}
\acmDOI{10.1145/3485983.3494850}
\acmISBN{978-1-4503-9098-9/21/12}

\begin{document}
\title[OnSlicing]{OnSlicing: Online End-to-End Network Slicing with Reinforcement Learning}

\author{Qiang Liu}
\affiliation{%
  \institution{University of Nebraska-Lincoln}
}
\email{qiang.liu@unl.edu}

\author{Nakjung Choi}
\affiliation{%
  \institution{Nokia Bell Labs}
}
\email{nakjung.choi@nokia-bell-labs.com}

\author{Tao Han}
\affiliation{%
  \institution{New Jersey Institute of Technology}
}
\email{tao.han@njit.edu}


\renewcommand{\shortauthors}{Q. Liu, et al.}

\begin{abstract}
Network slicing allows mobile network operators to virtualize infrastructures and provide customized slices for supporting various use cases with heterogeneous requirements. Online deep reinforcement learning (DRL) has shown promising potential in solving network problems and eliminating the simulation-to-reality discrepancy. Optimizing cross-domain resources with online DRL is, however, challenging, as the random exploration of DRL violates the service level agreement (SLA) of slices and resource constraints of infrastructures. In this paper, we propose OnSlicing, an online end-to-end network slicing system, to achieve minimal resource usage while satisfying slices' SLA. OnSlicing allows individualized learning for each slice and maintains its SLA by using a novel constraint-aware policy update method and proactive baseline switching mechanism. OnSlicing complies with resource constraints of infrastructures by using a unique design of action modification in slices and parameter coordination in infrastructures. OnSlicing further mitigates the poor performance of online learning during the early learning stage by offline imitating a rule-based solution. Besides, we design four new domain managers to enable dynamic resource configuration in radio access, transport, core, and edge networks, respectively, at a timescale of subseconds. We implement OnSlicing on an end-to-end slicing testbed designed based on OpenAirInterface with both 4G LTE and 5G NR, OpenDayLight SDN platform, and OpenAir-CN core network. The experimental results show that OnSlicing achieves 61.3\% usage reduction as compared to the rule-based solution and maintains nearly zero violation (0.06\%) throughout the online learning phase. As online learning is converged, OnSlicing reduces 12.5\% usage without any violations as compared to the state-of-the-art online DRL solution.
\end{abstract}



\maketitle

\vspace{-0.1in}
\section{Introduction}

The 5G scenarios, e.g., enhanced mobile broadband (eMBB), ultra-reliable low-latency communications (URLLC), and massive machine-type communication (mMTC), create new applications such as mobile augmented reality (MAR)~\cite{liu2018dare}, 360-degree video streaming, vehicle-to-everything (V2X), and Internet of things (IoT)~\cite{wp5d2017minimum}.
These emerging use cases have diverse requirements of quality of services (QoS), e.g., delay, jitters, throughput, and reliability. Hence, there is a pressing need for mobile network operators (MNOs) to customize the provisioning of communication, networking, and computing resources~\cite{afolabi2018network}. 
Network slicing enables these applications by virtualizing physical infrastructures such as base stations and switches, and providing logical networks (\emph{aka.} network slices) with dedicated virtual resources for slice tenants~\cite{foukas2017network}.
As the performances of slices are correlated to cross-domain network resources, an end-to-end slicing is necessary to create slices composed of resources in radio access networks (RAN), transport networks (TN), core networks (CN), and edge networks (EN).

Model-based methods model mobile networks with approximated mathematical models~\cite{salvat2018overbooking, d2020sl}, which cannot completely represent complicated network dynamics and thus fail to fulfill these distinct slice requirements.
Data-driven approaches, especially deep reinforcement learning (DRL), emerge in recent years~\cite{ayala2019vrain,niu2020billion,bega2019machine,liu2020edgeslice,yan2020learning} to tackle the high-dim correlations in complex networks.
These DRL solutions train their DRL policies within offline environments such as network simulators and apply the offline trained policy to control the real network directly.
In practice, these offline trained policies suffer from the discrepancy between simulated environments and real networks~\cite{mao2019learning, zhang2020onrl}.

Online DRL~\cite{gilad2020mpcc, zhang2020onrl} addresses this problem by allowing the DRL agent to directly learn from real networks.
For example, OnRL~\cite{zhang2020onrl} employs online DRL to learn the video streaming policy within real networks and optimize the video stalling rate.
However, the intrinsic exploration mechanism of online DRL, which explores a large action space containing all possible actions, can lead to violations of service-level agreements (SLAs).
It is necessary to maintain the performance requirement of slices throughout the online learning phase when managing real networks, where the random explored actions may result in severe performance degradation.

In this work, we propose OnSlicing, an end-to-end network slicing system, to enable online cross-domain resource orchestration with near-zero violations of slices' SLA throughout the online learning phase.
OnSlicing is accomplished with the following novel designs.

\textbf{Learning with near-zero violations.}
On managing an end-to-end network, MNO aims to minimize network resource usage while satisfying the performance requirement of slices.
The existing online DRL approaches~\cite{zhang2020onrl, gilad2020mpcc} rely on the random exploration mechanism to explore the action space and seek a better policy for improving the cumulative rewards without considering any system constraints.
As a result, the performance requirement of slices can be violated by either a sequence of undesired resource orchestration actions or a constraints-unaware policy update.
To resolve this issue, we design a \emph{constraint-aware policy update} method, in which the violation of slice's SLA is adaptively incorporated as a penalty into the reward function.
Besides, we design a \emph{proactive baseline switching} mechanism, in which the DRL policy is truncated in advance and a baseline policy (rule-based) is invoked for handling the rest of the episode if the predicted performance of the DRL policy does not meet the slice requirements.
In this way, OnSlicing can online learn the resource orchestration policy within real networks to minimize resource usage while maintaining near-zero violations of slices' SLA.

\textbf{Learning in distributed networks.}
The physical infrastructures, e.g., base stations, are geographically distributed in large-scale mobile networks.
The existing individualized DRL solutions~\cite{niu2020billion, zhang2020onrl,chu2020multi} generate actions independently and thus fail to meet resource constraints of infrastructures such as the total number of physical resource blocks (PRBs) in RAN.
To handle this challenge, we design a \emph{distributed coordination} mechanism that coordinates the resource orchestration between the individualized DRL agent of slices and infrastructures in a distributed manner.
In particular, we design an \emph{action modifier} in each DRL agent that modifies orchestration actions to satisfy the resource constraints while maintaining the instantaneous slice performance.
Moreover, we design a \emph{parameter coordinator} in each infrastructure to coordinate specific coordinating parameters with action modifiers in DRL agents.
In this way, OnSlicing can meet the resource constraints of infrastructures while maintaining the instantaneous performance requirement of slices.

\textbf{Learning from baseline.}
An online DRL policy usually performs worse than the rule-based policy at the early learning stage when an effective policy has not been learned yet~\cite{le2018hierarchical, saunders2017trial, ravichandar2020recent}.
Thus, allowing the DRL agent to manage real networks at the early stage could result in substantially poor performance and increase possibilities of SLA violation.
To address this issue, we design a \emph{learning from baseline} scheme, in which the DRL agent of slices are offline trained to imitate the behavior of the rule-based policy.
In this way, OnSlicing allows the DRL agent of slices to start online learning with a policy has a similar performance as the rule-based policy.
As a result, OnSlicing avoids excessive SLA violations at the early learning stage. 

\textbf{Domain managers.}
The foundation of end-to-end slicing~\cite{foukas2017network,alliance2016description} lies in the infrastructure virtualization in multiple technical domains.
We design four new domain managers in RAN, TN, CN, and EN, respectively, to efficiently virtualize physical infrastructures and assure performance isolation among slices.
In this way, OnSlicing can dynamically manage multiple end-to-end slices such as creation, deletion, and adjusting of a variety of network configurations at the timescale of subseconds.

\textbf{Contributions.}
To the best of our knowledge, OnSlicing is the first end-to-end network slicing system that minimizes resource usage with near-zero violations of slices' SLA in mobile networks.
The specific contributions of OnSlicing are summarized as follows:
\begin{itemize}[leftmargin=*]
    \item We design novel methods (Sec.~\ref{sec:safety_learn}) to apply online DRL to manage end-to-end slicing in real networks with near-zero violations of slices' SLA throughout the online learning phase.
    \item We design novel mechanisms (Sec.~\ref{sec:scalable_learn}) to allow the individualized DRL agent of slices to meet resource constraints in infrastructures, and develop a new learning scheme (Sec.~\ref{sec:imitate_learn}) to mitigate the poor performance of online learning during the early learning stage. 
    \item We develop four domain managers (Sec.~\ref{sec:implementation}) in RAN, TN, CN, and EN, which enable dynamic resource configurations at the timescale of subseconds.
    \item We implement OnSlicing on an end-to-end slicing testbed (Sec.~\ref{sec:implementation}) and validate that OnSlicing significantly outperforms state-of-the-art solutions in terms of resource usage and SLA violation (Sec.~\ref{sec:evaluation}).
\end{itemize}







\begin{figure}[!t]
	\centering
	\includegraphics[width=3.0in]{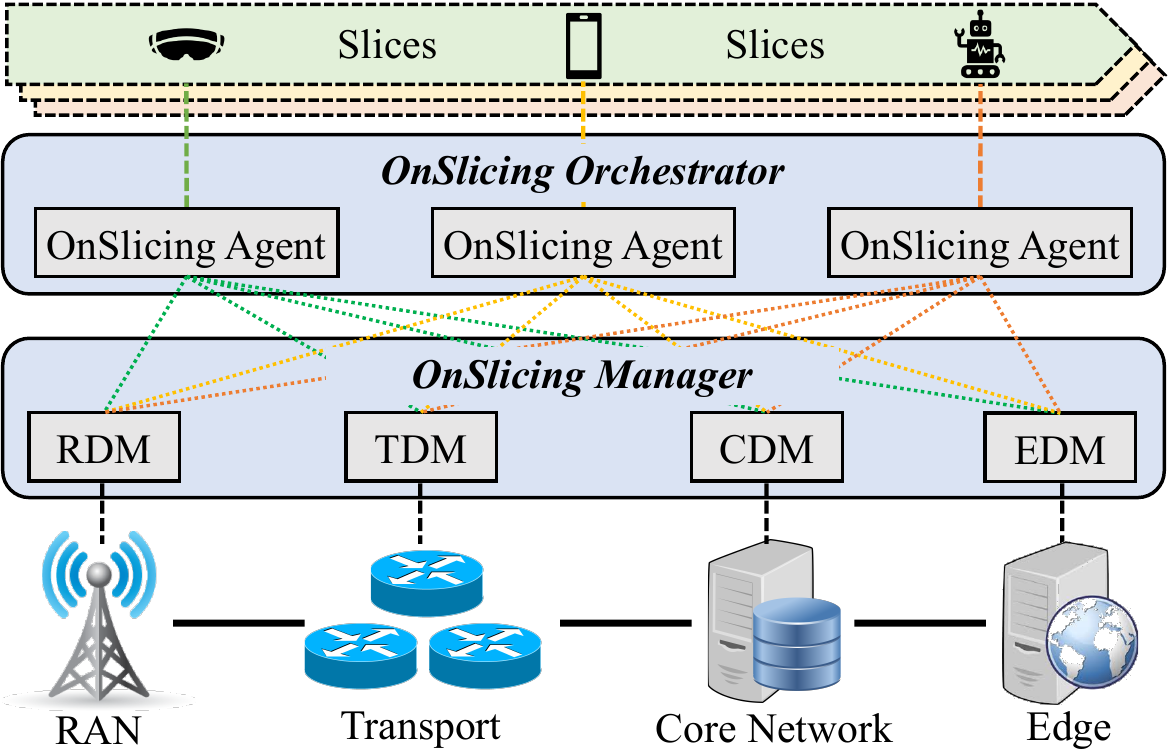}
	\caption{\small The OnSlicing system.}
	\label{fig:system_overview}
\end{figure}

\section{System Overview}

\textbf{End-to-End Slicing.}
An end-to-end network slice refers to a virtual network with a collection of all needed network resources to meet the performance of particular services or applications operated by slice tenants~\cite{alliance2016description}.
For example, an end-to-end slice tailored for mobile AR/VR may include radio transmission resources in RAN, data transportation resources in TN, packet processing resources in CN, and computation and storage resources in EN.
As a slice tenant creating its network slice, it makes a service level agreement (SLA) with MNO, which specifies the performance requirement, e.g., delay, throughput, and reliability.

To accomplish end-to-end slicing, MNO needs to provide two essential attributes~\cite{foukas2017network}, i.e., performance isolation that assure the performance of a slice is not influenced by any operations of the other slices, and SLA assurance that satisfies the performance requirement of slices.
Meanwhile, MNO aims to serve these slices with the minimum resource usage~\cite{salvat2018overbooking} and the total resources are constrained by physical infrastructures.

\textbf{Overview.}
As shown in Fig.~\ref{fig:system_overview}, OnSlicing consists of two main components, i.e., the manager and the orchestrator.

The OnSlicing manager virtualizes end-to-end infrastructures to virtual resources and implements resource orchestration actions at runtime.
It consists of a radio domain manager (RDM), transport domain manager (TDM), core domain manager (CDM), and edge domain manager (EDM) for managing eNBs/gNBs in RAN, switches/routers in TN, virtual network functions (VNFs) in CN, and servers in EN, respectively.
We design these domain managers to assure the performance isolation among slices and achieve low-overhead virtualization.

The OnSlicing orchestrator allocates end-to-end virtual resources to various slices and satisfies their diverse performance requirements.
It includes multiple OnSlicing agents, where each agent allocates virtual resources for a slice.
We design these agents to online learn from the interactions with domain managers in real networks, and maintain the slices' SLA and resource constraints in infrastructures throughout the online learning phase.

\section{Online Learning with near-zero Violations}
\label{sec:safety_learn}

\begin{figure*}[!t]
	\centering
	\includegraphics[width=6.7in]{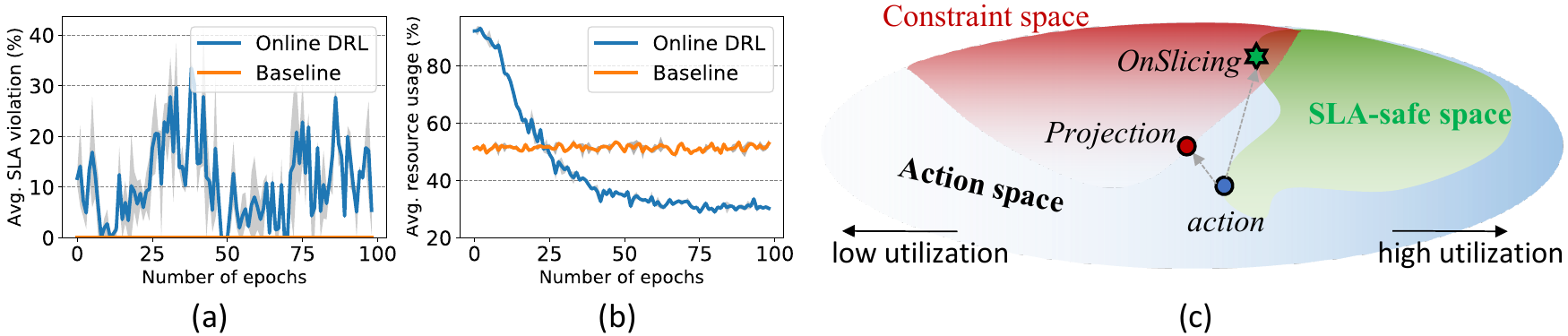}
	\caption{\small a) Avg. SLA violation under existing solutions. b) Avg. resource usage under existing solutions. c) An illustration of action modification.}
	\label{fig:motivation}
\end{figure*}

\textbf{Unsafe DRL exploration.}
Existing DRL approaches~\cite{mnih2013playing, mnih2015human} aim to maximize the cumulative rewards without considering practical constraints in networks.
The performance requirement of slices could be easily violated when applying these approaches to minimize resource usage.
Furthermore, the orchestration actions generated by a DRL policy could result in poor performance of slices, due to the intrinsic exploration of DRL algorithms that randomly explores different actions in the whole action space for a better reward.
In Fig.~\ref{fig:motivation} (a), we show the average SLA violation of slices under a baseline policy and a DRL agent whose reward function is penalized with a fixed weight if the slice SLA is violated.
We observe the DRL agent could have more than 30\% violation of the slice SLA during the online learning phase, while the baseline policy has no violation.
Thus, it is unsafe to allow the DRL agent to online learn from real networks without any safety mechanisms.

\begin{figure}[!t]
	\centering
	\includegraphics[width=3.33in]{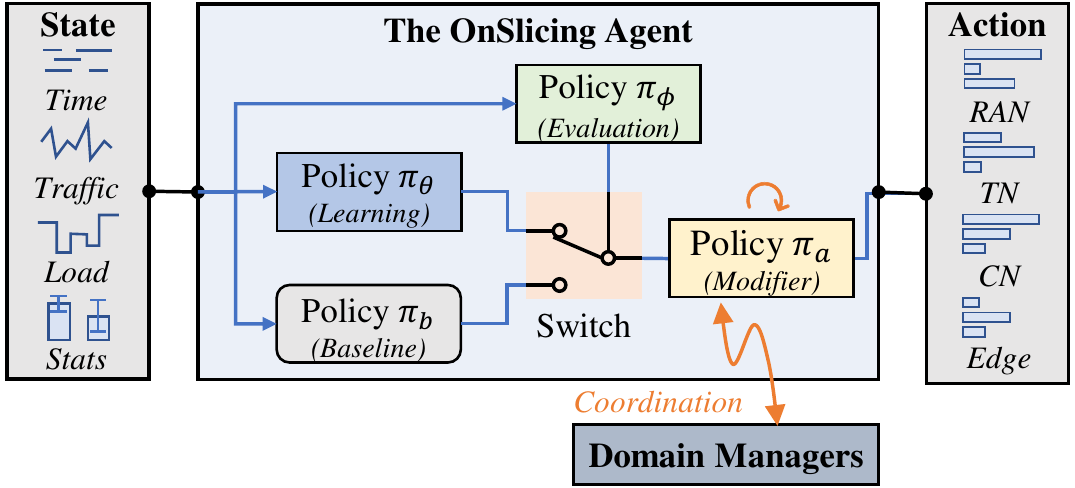}
	\caption{\small The orchestration agent.}
	\label{fig:alg_overview}
\end{figure}

\textbf{OnSlicing approach.}
OnSlicing applies online DRL to orchestrate cross-domain resources while maintaining the performance requirement of slices as shown in Fig.~\ref{fig:alg_overview}.
First, instead of centrally managing all slices, we design an OnSlicing agent for each slice,  which is more efficient to scale in dynamic network slicing and incremental network deployment.
Second, we design a constraint-aware policy update method to adaptively incorporate the violation of slices' SLA into the reward function as a penalty, which updates policy $\pi_\theta$ to avoid the actions that violate the slice SLA.
Third, we design a proactive baseline switching mechanism to switch to the baseline policy $\pi_b$ for managing resource orchestration if policy $\pi_\theta$ is predicted to violate the slice SLA.
Fourth, we realize the policy update method based on the state-of-the-art proximal policy optimization (PPO)~\cite{schulman2017proximal} to ensure a smooth performance improvement and prevent excessive policy update steps.

\textbf{The problem.}
We aim to derive an optimal policy $\pi^*_\theta$ that minimizes the usage of virtual resources without violating the slice SLA.
Therefore, given a time period $\mathcal{T}$, e.g., 24 hours, we formulate the resource orchestration problem $\mathbb{P}_0$ as
\begin{align}
     \mathbb{P}_0: \max \limits_{ \pi_\theta } & \;\;\;\;\; \mathbb{E}_{\pi_\theta} \left[ \sum\nolimits_{t \in \mathcal{T}} {r(\mathbf{s}_t, \mathbf{a}_t)} \right]  \\ 
     s.t. &\;\;\;\;\; \mathbb{E}_{\pi_\theta} \left[ \frac{1}{T} \sum\nolimits_{t \in \mathcal{T}} c(\mathbf{s}_t, \mathbf{a}_t) \right] \le {C}_{\max}, \label{const:safe}
\end{align}
where $\mathbf{s}_t$, $\mathbf{a}_t$ and $r(\mathbf{s}_t, \mathbf{a}_t)$ are the network state, orchestration action and reward of the slice at time slot $t$.
The resource orchestration actions are made at the beginning of every time slot.
The constraint in Eq.~\ref{const:safe} ensures the statistical performance of the slice is met, where $c(\mathbf{s}_t, \mathbf{a}_t)$ and $C_{\max}$ are the cost and SLA threshold, respectively.

\textbf{Constraints-Aware Update.}
To make policy $\pi_\theta$ aware of the slice SLA, we use the Lagrangian primal-dual method~\cite{boyd2004convex} to incorporate the constraints into the reward function.
Specifically, we build Lagrangian as 
\begin{align}
    \mathcal{L} &= \mathbb{E}_{\pi_\theta} \left[ \sum\nolimits_{t \in \mathcal{T}} { \left(r(\mathbf{s}_t, \mathbf{a}_t) - \frac{\lambda}{T}  c(\mathbf{s}_t, \mathbf{a}_t) \right)} \right]  + \lambda {C}_{\max}, \label{eq:lagrangian}
\end{align}
where $\lambda$ is the multiplier.
The problem can be addressed by alternatively solving the primal problem expressed as 
\begin{equation}
\pi^*_\theta = \arg\max\limits_{\pi_\theta}  \mathcal{L}(\pi_\theta, \mathbf{\lambda}),
\label{eq:primal_problem}
\end{equation} 
and the dual problem $\mathbf{\lambda}^* = \arg\min \limits_{ \mathbf{\lambda} \ge 0} \mathcal{L}(\pi_\theta, \mathbf{\lambda})$.
The dual problem is solved by updating the multiplier with the sub-gradient descent~\cite{boyd2004convex} as follows 
\begin{align}
    \lambda = \left[\lambda + \varepsilon \left( \mathbb{E}_{\pi_\theta} \left[ \frac{1}{T} \sum\nolimits_{t \in \mathcal{T}} c(\mathbf{s}_t, \mathbf{a}_t) \right] - {C}_{\max} \right)\right]^+,
    \label{eq:update_multiplier}
\end{align}
where $[x]^+ = max(x,0)$ and $\varepsilon$ is the step size.
In this way, the multiplier $\lambda$ is increased if the performance requirement of the slice is violated.

\textbf{Proactive Baseline Switching.}
Although the statistical performance requirement (Eq.~\ref{const:safe}) of the slice could be satisfied eventually using the constraint-aware policy update method, we find that the slice SLA could be violated during the online learning phase due to the intrinsic DRL exploration.
To mitigate the SLA violation, we design a proactive baseline switching mechanism.
The fundamental idea is to \emph{let the baseline policy take over the rest of the episode if the cumulative cost at the current time slot plus the expected cost value function of the baseline policy is larger than the SLA threshold ${C}_{\max}$}\footnote{The estimation of the cost value function and the switching decision are made at every time slot.}.
The cost value function is defined as $C = \mathbb{E}_{\pi_b} \left[\sum\nolimits_{t=t_c}^{\mathcal{T}}{c(\mathbf{s}_t, \pi_b(\mathbf{s}_t))} \right]$, in other words, the cumulative cost starts at current time slot $t_c$ if we follow the baseline policy $\pi_b$ until the end of the episode.

The cost value function correlates to the high-dim network state, which is very complicated and can not be mathematically represented.
Thus, we create a neural network with policy $\pi_\phi$ to learn the cost value function of the baseline policy under different states.
Although deterministic neural networks can be used to estimate the cost value, they only generate a single estimation value and overlook statistical information.
For example, if the cost value has a small mean value but a large deviation, switching to the baseline merely based on the mean value could be too late and thus result in a large probability of SLA violation.
Therefore, we leverage the variational inference technique~\cite{hoffman2013stochastic, zhang2018advances} to learn its probabilistic model, e.g., mean and deviation.

The variational inference approximates unknown complex distributions with a cluster of tractable distributions, e.g., Gaussian distributions.
Here, the posterior distribution $p(\phi|\mathcal{D})$, where $\mathcal{D}$ are the observed cost values, is approximated by minimizing the KL-divergence $D_{KL} \left[ q(\phi) || p(\phi|\mathcal{D}) \right]$, where $q(\phi)$ is the Gaussian distribution.
Although the posterior distribution is unable to calculate, the minimization of KL-divergence~\cite{kullback1997information} corresponds to the maximization of the evidence lower bound (ELBO)~\cite{saul1996mean} based on the following equation 
\begin{align}
    \log p(\mathcal{D}) = \underbrace{ \underset{q(\phi)}{\mathbb{E}}{\log\frac{p(\phi, \mathcal{D})}{q(\phi)}} }_{\mathbf{ELBO}}  +  D_{KL} \left[ q(\phi) || p(\phi|\mathcal{D}) \right]. 
\end{align}
Since the KL-divergence is always positive and $\log p(\mathcal{D})$ is irrelevant to $\phi$, we can train policy $\pi_\phi$ by maximizing the ELBO rewritten as 
\begin{align}
    ELBO  =  \underset{q(\phi)}{\mathbb{E}}{\log p(\mathcal{D} |\phi)}  -  D_{KL} \left[ q(\phi) || p(\phi) \right].
    \label{eq:elbo} 
\end{align}
On the right side of Eq.~\ref{eq:elbo}, the first part is the likelihood, and the second part is the KL-divergence between the priors. By assuming that the priors are Gaussian distributions, both parts are tractable and can be calculated effectively.

Therefore, we can use policy $\pi_\phi$ to estimate the cost value function under state $\mathbf{s}_t$ and obtain the mean $\mu$ and standard deviation $\sigma$. 
Then, we determine the policy switching between policy $\pi_\theta$ and the baseline policy $\pi_b$ at time slot $t$ according to 
\begin{align}
    \mathbf{a}_t = \left\{\begin{matrix}
    \pi_b(\mathbf{s}_t), & E_t \geq T \cdot {C}_{\max},\\ 
    \pi_\theta(\mathbf{s}_t), & E_t < T \cdot  {C}_{\max},
    \end{matrix}\right.
    \label{eq:switching_criteria} 
\end{align}
where $E_t = \sum\nolimits_{m = 0 }^{t} c(\mathbf{s}_m, \mathbf{a}_m) + \mu + \eta \cdot \sigma$, and $\eta$ is a factor to control risk preferences. When $\eta$ is high, the switching decision is more sensitive to the deviation of the cost value, i.e., the baseline policy might be invoked earlier, and thus this baseline switching mechanism becomes more conservative.

\textbf{Smooth Policy Improvement.}
\label{sec:agent_design}
We develop the training method for policy $\pi_\theta$, which corresponds to solving the primal problem in Eq.~\ref{eq:primal_problem}~\cite{paternain2019constrained}.
We train policy $\pi_\theta$ based on the state-of-the-art PPO algorithm~\cite{schulman2017proximal} rather than deep deterministic policy gradient (DDPG) algorithm~\cite{lillicrap2015continuous}.
Because the DDPG algorithm improves policies by implicitly minimizing the mean square Bellman error (MSBE)~\cite{lillicrap2015continuous}, which tends to have changing performance in practice~\cite{duan2016benchmarking} and thus leads to excessive switching to the baseline policy.
In contrast, the PPO algorithm directly maximizes the expected return and enables smooth performance improvement by using a clipped surrogate objective to prevent too large policy update steps.

With the proactive baseline switching mechanism, an episode of resource orchestration actions could be composed of both policy $\pi_\theta$ and the baseline policy $\pi_b$.
In this situation, updating policy $\pi_\theta$ with the whole mixed episode diverges the training of the DRL agent, because the partial episode run by the baseline policy does not apply to policy $\pi_\theta$.
To address this issue, we only use the effective transitions run by policy $\pi_\theta$ and discard the remaining episode run by the baseline policy.
Meanwhile, we estimate the reward value function $R =  \mathbb{E}_{\pi_\theta} \left[ \sum\nolimits_{t=t_r}^{\mathcal{T}}{r(\mathbf{s}_t, \pi_\theta(\mathbf{s}_t)} \right]$ at the truncated time slot $t_r$, which helps in calculating accurate reward value function of truncated episodes.

\emph{\textbf{State}}:
We define the state space as the combination of the current time slot $t$, the traffic of slice $f_{t-1}$, the average channel condition of slice users $h_{t-1}$, the average radio resource usage in RAN $g_{t-1}$, the average workload of VNFs and edge server $w_{t-1}$, the last slice performance and cost $r_{t-1}, c_{t-1}$, the slice SLA threshold $C_{\max}$ and cumulative cost at current time slot $\sum\nolimits_{m=0}^t c(\mathbf{s}_m, \mathbf{a}_m)$.
The state is designed to reveal the informative slice statistics and comprehensive network status to policy networks in the OnSlicing agent.
In the state space, $[t, f_{t-1}]$ provide the information about the expected traffic at time slot $t$, $[h_{t-1}, g_{t-1}, w_{t-1}]$ suggest the network status about the resource usage,  $[r_{t-1}, c_{t-1}]$ indicate the potential lasting influence from time slot $t-1$, and $[C_{\max}, \sum\nolimits_{m=0}^t c(\mathbf{s}_m, \mathbf{a}_m)]$ show the slice status about the performance requirement.

\emph{\textbf{Action}}:
The action is designed to allow the OnSlicing agent to orchestrate virtual resources to the slice in different domains.
We identify multiple key factors that could affect the performance of slices (Sec.~\ref{sec:implementation}).
We define the action space as the combination of uplink radio bandwidth $U_u$, uplink MCS offset $U_m$, uplink scheduling algorithm $U_a$, downlink radio bandwidth $U_d$, downlink MCS offset $U_s$, downlink scheduling algorithm $U_g$, transport bandwidth $U_b$ and reserved path in TN $U_l$, CPU allocation $U_c$ and RAM $U_r$ allocation for co-located SGPW-U and edge server. 

\emph{\textbf{Reward}}:
We define the reward function as the negative total virtual resource usage of the slice 
\begin{equation}
r(\mathbf{s}_t, \mathbf{a}_t) = - \left( U_u + U_d + U_b + U_l + U_c + U_r \right). 
\label{eq:reward_calculation} 
\end{equation}
Here, without loss of generality, we sum up all the used virtual resources by using the same weights.
The scheduling algorithm and MCS offset in both uplink and downlink are not counted in the reward function because their selections implicitly impact the radio resource usage.

\emph{\textbf{Cost}}:
The cost indicates how much performance degradation the slice experienced as compared to its performance requirement. The cost is reported by the slice every time slot.
Without loss of generality, we define the cost function as 
\begin{equation}
c(\mathbf{s}_t, \mathbf{a}_t) = 1 - clip\left(p_t(\mathbf{s}_t, \mathbf{a}_t)/P, 0, 1\right), 
\end{equation}
where $P$ is the performance requirement of the slice, and $clip(x,y,z)$ means clipping $x$ between $y$ and $z$. 
For example, a video streaming slice needs an FPS $P=30$, then a cost 0.33 can be observed if $p_t=20$ at the current time slot.


\section{Individualized Learning in Distributed Networks}
\label{sec:scalable_learn}

\textbf{Constraints in distributed infrastructures.}
The virtual resources can be limited by physical infrastructures, e.g., the total number of PRBs in RAN and the computing capacity in edge servers.
As every OnSlicing agent generates orchestration actions according to its network state independently, some resources could be over-requested, i.e., the total requested resources exceeds the resource capacity.
As shown in Fig.~\ref{fig:motivation} (c), an action is generated by the OnSlicing agent with the minimum resource usage in the SLA safe space.
The actions in the SLA safe space can meet the performance requirement of the slice, and the actions in the constraint space can satisfy the resource capacity.
The existing method requires domain managers to scale down all actions of slices, i.e., projection, if the summation of requested resources surpluses the capacity of the infrastructure.
However, we find this method leads to substantial performance degradation and possible SLA violation in long-term because the requested resources of slices are under-provisioned.
Thus, it is needed to modify the orchestration action of slices within the SLA safe space while maintaining their instantaneous performances, i.e., in the constraint space.

\textbf{OnSlicing approach.}
OnSlicing satisfies the resource constraints and keeps the performance of slices, as shown in Fig.~\ref{fig:alg_overview}, by designing a distributed coordination mechanism.
On the one hand, we design an action modifier in each OnSlicing agent to modify the original action generated by policy $\pi_\theta$ according to the parameters from domain managers. The modified action maintains the slice performance with the minimum distance to the original action.
On the other hand, we design a parameter coordinator in each domain manager to adaptively adjust the coordinating parameters exchanged with action modifiers and coordinate the resource usage among slices.
Furthermore, we define the initialization method of the coordinating parameters at every time slot to reduce the needed number of interactions between OnSlicing agents and domain managers.

\textbf{The problem.}
We aim to find modified actions for all slices denoted as $\hat{\mathbf{a}}^i_t, \forall i \in \mathcal{I}$.
The modified actions need to reduce their distances to original actions $\mathbf{a}^i_t, \forall i \in \mathcal{I}$ and the cost of slices while maintaining the resource constraints.
On the one hand, we need to stick to the original action generated by policy $\pi_\theta$, which could achieve the minimum long-term resource usage.
On the other hand, we need to reduce the instantaneous cost of slices, which helps maintain the slice SLA in the long-term. 
Thus, we formulate the action modification problem as 
\begin{align}
     \min \limits_{ \hat{\mathbf{a}}^i } \;\;& \sum\nolimits_{i \in \mathcal{I}} \left\{ |\hat{\mathbf{a}}_t^i - \mathbf{a}_t^i|^2_2  + c(\mathbf{s}^i_t, \hat{\mathbf{a}}_t^i) \right\} \\
     s.t. \;\;& \sum\nolimits_{i \in \mathcal{I}}{\hat{\mathbf{a}}^{i,k}_t } \leq L_{\max}^{k}, \forall k \in \mathcal{K},
     \label{prob:scalablility} 
\end{align}
where the superscript $i$ denotes the slice $i$, $L_{\max}^{k}$ is the capacity of the $k$th resource and $|\cdot|_2^2$ is the $l_2$-norm operation.

This problem involves the resource orchestration of all slices and resource capacity constraints in all domain managers in the distributed network.
A centralized approach could lead to excessive communication overhead and delay between OnSlicing agents and domain managers.

\textbf{Action modification.}
We create an action modifier in each OnSlicing agent to modify the action generated by policy $\pi_\theta$ and maintain resource constraints in different domain managers.
Specifically, the action modifier generates an action $\hat{\mathbf{a}}_*^i$ to minimize the following objective function 
\begin{align}
     \hat{\mathbf{a}}_*^i = \arg\min \limits_{ \hat{\mathbf{a}}^i_t } \{ \underbrace{|\hat{\mathbf{a}}_t^i - \mathbf{a}_t^i|^2_2  + \sum\nolimits_{k \in \mathcal{K}}{\beta^k_t \hat{\mathbf{a}}^{i,k}_t } + c(\mathbf{s}_t^i, \hat{\mathbf{a}}_t^i)}_{\textbf{H}_t} \},
     \label{prob:action_modify} 
\end{align}
where the resource constraints in Eq.~\ref{prob:scalablility} are incorporated into the objective function with $\beta^k_t, \forall k \in \mathcal{K}$.
Here, we define $\beta^k_t, \forall k \in \mathcal{K}$, as the coordinating parameters for regulating the resource orchestration in OnSlicing agents, which are updated in domain managers.

Although $\mathbf{a}_t^i$ and $\beta^k_t, \forall k \in \mathcal{K}$ in Eq.~\ref{prob:action_modify} are known, the cost function of the slice, i.e., $c(\mathbf{s}_t^i, \hat{\mathbf{a}}_t^i), \forall i \in \mathcal{I}$, is too complicated to be mathematically modeled.
To this end, we design a neural network with policy $\pi_a$ in the action modifier to solve the problem in Eq.~\ref{prob:action_modify} and generate the modified action $\hat{\mathbf{a}}_t^i$.
The inputs of policy $\pi_a$ are the combination of current state $\mathbf{s}_t^i$, original action $\mathbf{a}_t^i$ generated by policy $\pi_\theta$, and coordinating parameters $\beta^k_t, \forall k \in \mathcal{K}$.
This network is offline trained with supervised learning by minimizing the objective function in Eq.~\ref{prob:action_modify}.
The training dataset includes the pairs of $[\mathbf{s}_t^i, \mathbf{a}_t^i, \beta^k_t, \forall k \in \mathcal{K}]$ and $\textbf{H}_t$.
We build the dataset by collecting state-action-cost pairs $[\mathbf{s}_t^i, \mathbf{a}_t^i, c(\mathbf{s}_t^i, \hat{\mathbf{a}}_t^i)]$ from the real system, appending randomly generated coordinating parameters to each state-action pair, and calculating the objective function $\textbf{H}_t$.

\textbf{Parameter coordination.}
To comply with the resources constraints in Eq.~\ref{prob:scalablility}, we design a parameter coordinator in each domain manager for updating the coordinating parameters $\beta^k_t, \forall k \in \mathcal{K}$.
The coordinator of domain manager $k$ updates the coordinating parameters by using the sub-gradient descent method~\cite{boyd2004convex} as 
\begin{equation}
    \beta^k =  \left[\beta^k + \epsilon (\sum\nolimits_{i \in \mathcal{I}} \hat{\mathbf{a}}^{i,k}_t - {L}^k_{\max})\right]^+, 
\end{equation}
where $[x]^+ = max(x,0)$ and $\epsilon$ is a positive step size.
Here, the coordinating parameters $\beta^k$ are increased when the resource in this domain manager is over-requested, which consequently guides policy $\pi_a$ in the action modifier of OnSlicing agents.

In this way, the action modifier in OnSlicing agents exchange coordinating parameters $\beta^k, \forall k \in \mathcal{K}$ and coordinate the resource usage with domain managers until resource constraints are met.
However, the number of interactions between them could be large if the coordinating parameters $\beta^k, \forall k \in \mathcal{K}$ are initialized for every time slot.
To accelerate the convergence of the interactions, we use the coordinating parameters $\beta^k_{t-1}, \forall k \in \mathcal{K}$ at the last time slot as the start point $\beta^k_{t}, \forall k \in \mathcal{K}$ at current time slot.

\section{Offline Learning from Baseline}
\label{sec:imitate_learn}

\textbf{Poor policy at early stage.}
It is inefficient to allow a DRL agent to online learn from scratch within real networks because the agent usually requires a large number of training steps to obtain a policy with acceptable performances.
As a result, the DRL agent does not perform well, usually even worse than the baseline policy, at the early learning stage.
In Fig.~\ref{fig:motivation} (a) and (b), we show the online learning performance of the DRL agent and the baseline policy.
Although the DRL agent could explore a better policy than the baseline policy eventually, it is outperformed by the baseline policy in both resource usage and SLA violation at the early stage, i.e., until epoch 20.

\begin{figure}[!t]
	\centering
	\includegraphics[width=3.1in]{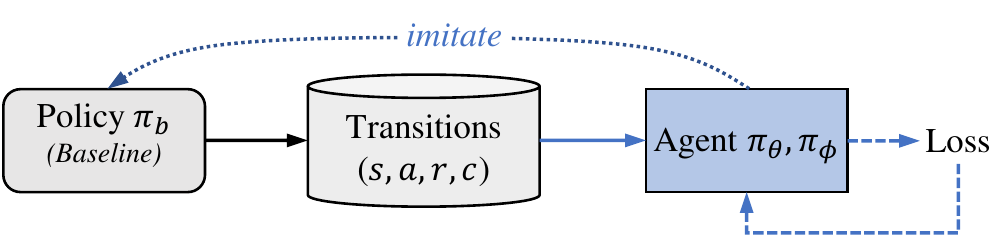}
	\caption{\small Imitation from the baseline policy}
	\label{fig:imitate_learning}
\end{figure}

\textbf{OnSlicing approach.}
We design OnSlicing agents, as shown in Fig.~\ref{fig:imitate_learning}, to offline imitate the baseline policy based on the dataset collected from the interactions between the baseline policy and real networks.
On the one hand, we train policy $\pi_\theta$ to imitate the baseline policy, i.e., taking similar actions as the baseline policy does, and to achieve similar performances such as resource usage and SLA violation. 
On the other hand, we train policy $\pi_\phi$ to estimate the cost value function of the baseline policy, which is used to determine the proactive baseline switching in Sec.~\ref{sec:safety_learn}. 
In this way, OnSlicing agents start their online learning in real networks with a similar performance as the baseline policy.

\textbf{Behavior cloning.}
We train policy $\pi_\theta$ based on behavior cloning (BC) to minimize the differences of generated actions by policy $\pi_\theta$ and the baseline policy $\pi_b$ with supervised learning.
Specifically, we collect the transitions, e.g., state-action pairs, of the baseline policy when it interacts with real networks.
Then, we train policy $\pi_\theta$ by minimizing the loss function  
\begin{equation}
    Loss = \frac{1}{|\mathcal{B}|}\sum\limits_{n \in \mathcal{B}}|\pi_b (\mathbf{s}_n) - \pi_\theta (\mathbf{s}_n)|_2^2,
    \label{eq:imitate_learning} 
\end{equation}
where $\mathbf{s}_n$ and $\mathcal{B}$ are the sampled state and batch of transitions, respectively.

In addition, we offline train policy $\pi_\phi$ to predict the cost value function of the baseline policy.
Specifically, we first collect states $\mathbf{s}_t$ and costs $ c(\mathbf{s}^i_t, \hat{\mathbf{a}}_t^i)$ run by the baseline policy $\pi_b$, and calculate the cost value function under different states.
The policy $\pi_\phi$ is then updated by maximizing the ELBO in Eq.~\ref{eq:elbo}.
To adapt to new states appearing during the online learning phase, policy $\pi_\phi$ is also updated as more transitions are observed.

\section{System Implementation}
\label{sec:implementation}
\begin{figure}[!t]
	\centering
	\includegraphics[width=3.33in]{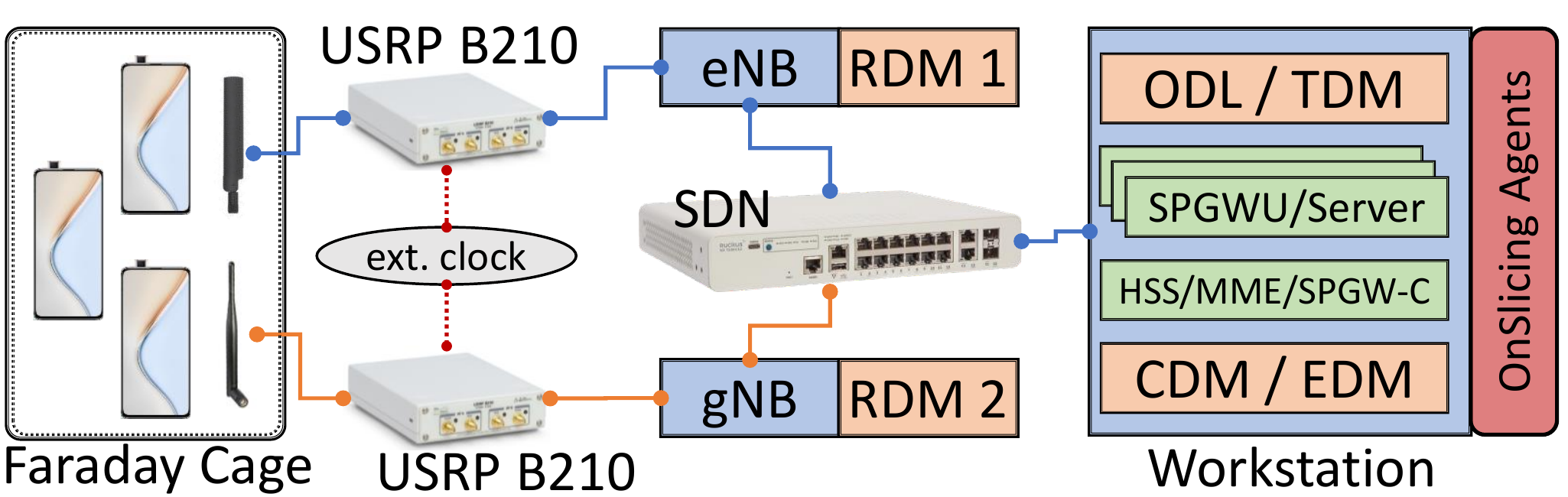}
	\caption{\small The OnSlicing testbed.}
	\label{fig:testbed}
\end{figure}

In this section, we present the OnSlicing implementation, including the domain managers and the hardware details of the testbed shown in Fig.~\ref{fig:testbed}.
The domain managers are developed to virtualize physical infrastructures in RAN, TN, CN, and EN, into virtual resources, respectively, and execute the resource orchestration actions generated by OnSlicing agents.
The design goal is to reduce the virtualization overheads while maintaining the performance isolation among slices.
We create a unified interface based on the REST API~\cite{masse2011rest} to facilitate the interactions between OnSlicing agents and domain managers.

\begin{figure}[!t] 
\captionsetup{justification=centering}
  \begin{minipage}[t]{0.235\textwidth}
    \centering
    \includegraphics[width=1.65in]{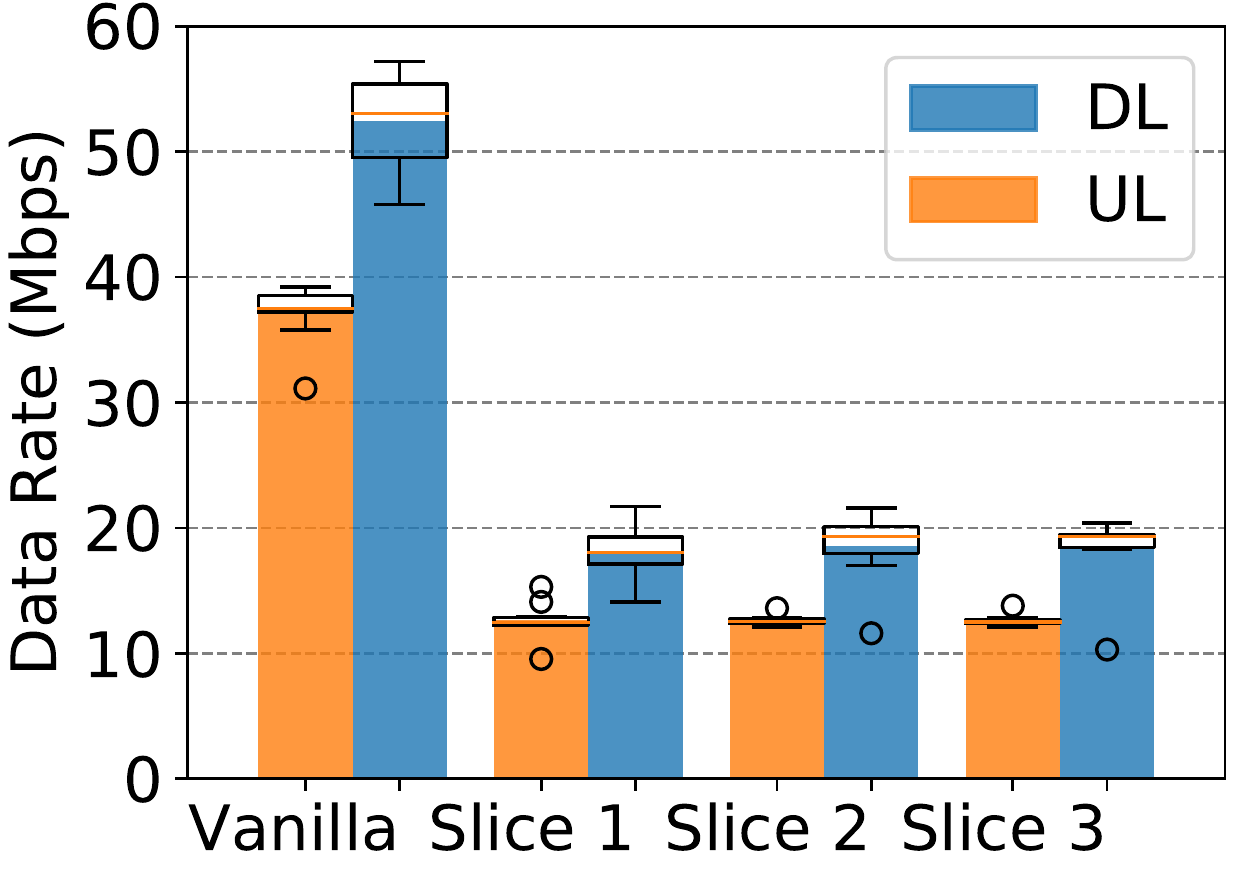}
    \captionof{figure}{\small The data rate of slices with RDM}
    \label{fig:virtualization_an_bw}
  \end{minipage}
  \hfill
  \begin{minipage}[t]{0.235\textwidth}
    \centering
    \includegraphics[width=1.65in]{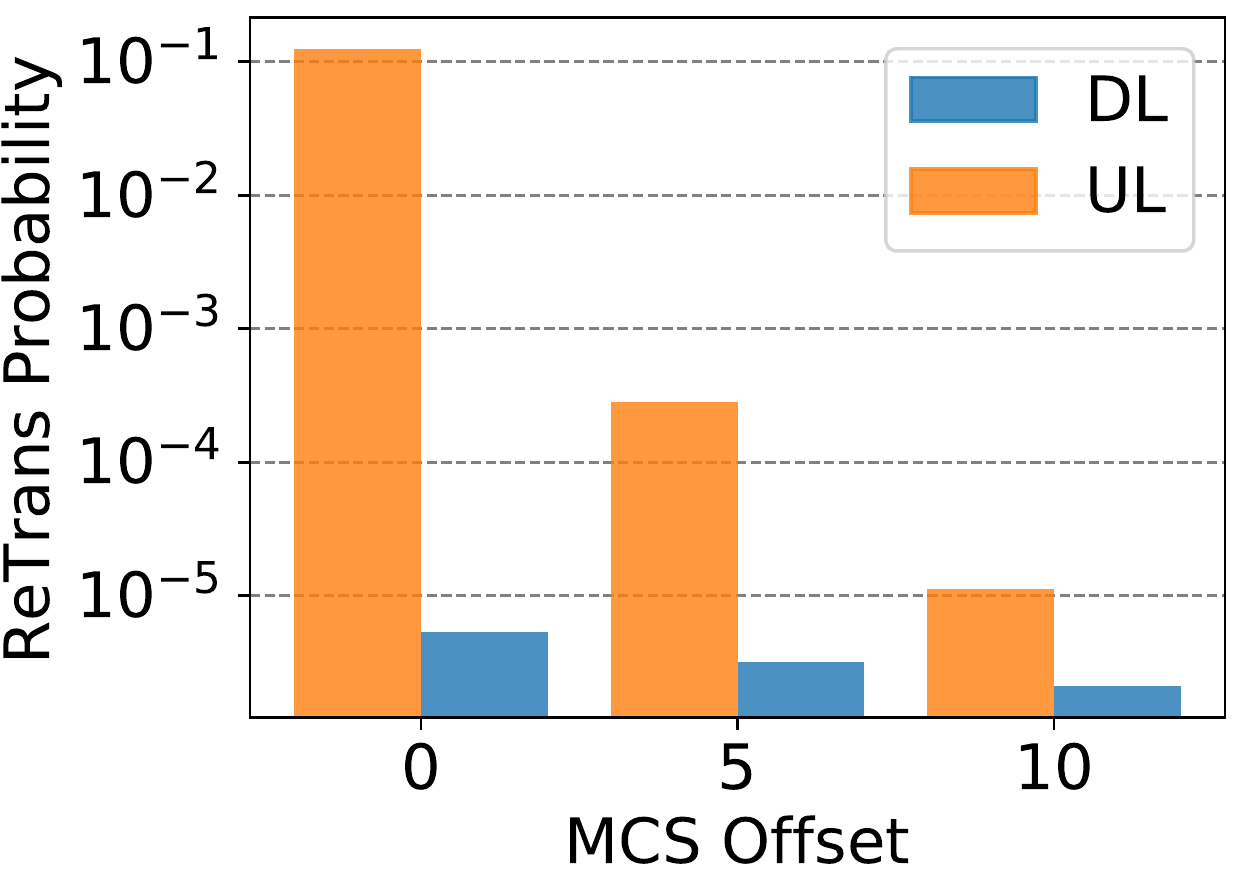}
    \captionof{figure}{\small MCS offset vs. Retransmission}
    \label{fig:virtualization_an_reliability}
  \end{minipage}
\end{figure}

\textbf{Radio domain manager.}
We design the radio domain manager (RDM) to slice 4G LTE and 5G NR RAN with customized CQI-MCS mapping tables for different slices.
The performance isolation among slices is guaranteed by exclusively assigning resource block groups (RBGs) and physical resource blocks (PRBs) in the downlink and uplink MAC layers, respectively.
In Fig.~\ref{fig:virtualization_an_bw}, we show the measured data rate of different slices that are assigned by the same virtual radio resources.
It can be seen that the total data rate of all slices nearly equals that of the vanilla system (OAI~\cite{OAI}), which verifies the low-overhead virtualization of RDM.

Besides, we introduce a new customized CQI-MCS mapping table for different slices to further improve the link reliability of the radio transmission.
Specifically, a slice can request an MCS offset in advance to counter the channel dynamics\footnote{The bit error rate (BER) is reduced if adopting a lower modulation scheme under the same power allocation~\cite{goldsmith2005wireless}.}. 
The used MCS by the slice is the vanilla MCS derived from the current CQI minus the MCS offset.
For example, a uRLLC slice can map CQI index 15 to 16-QAM instead of standardized 64-QAM to achieve more robust radio transmissions but lower link capacities.
In Fig.~\ref{fig:virtualization_an_reliability}, we show the re-transmission probability under different MCS offsets, which is calculated by the number of re-transmission PRBs over the total used number of PRBs for a slice using the \emph{iperf} tool.
It can be observed that the larger MCS offset the slice assigned, the lower the re-transmission probability the slice achieved, especially for the uplink transmission.

We develop the RDM based on OpenAirInterface (OAI)~\cite{OAI} with FlexRAN~\cite{foukas2016flexran}.
We use two Intel i7 computers to run the eNB and gNB that operate at 2.6 GHz (20MHz) and 3.5 GHz (40MHz), respectively.
Each computer is with a low-latency kernel of Ubuntu 18.04 and an Ettus USRP B210 as the RF front-end.
We use three 5G smartphones (POCO F2 Pro) that support both LTE and 5G NSA (EN-DC) as mobile users.
To eliminate external radio interferences, we use a Faraday cage to contain smartphones and antennas of eNB and gNB.
An Ettus Octo-clock is used to provide external 10MHz reference signals for both eNB and gNB.

\begin{figure}[!t]
	\centering
	\includegraphics[width=3.0in]{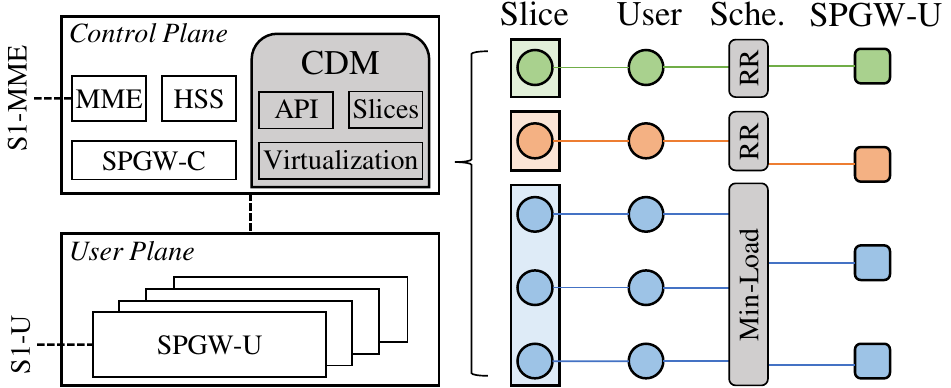}
	\caption{\small The CN slicing with core domain manager}
	\label{fig:cn_slicing}
\end{figure}

\textbf{Core domain manager.}
We design the core domain manager (CDM), as shown in Fig.~\ref{fig:cn_slicing}, to enable an isolated user plane for each slice, i.e., SPGW-U in EPC, by leveraging the CUPS-based core network architecture.
Each slice is associated with a set of SPGW-U instances and a corresponding SPGW-U scheduling method.
The method determines the associated slice users based on their international mobile subscriber identities (IMSIs).
It selects the destination SPGW-U from the SPGW-U pool of the slice based on the round-robin scheduling during the initial attachment procedure of users. 
We exclusively associate an SPGW-U instance to a slice, which ensures performance isolation.
We develop the CDM based on OpenAir-CN~\cite{openaircn} and deploy it on the workstation computer with an Intel i7 CPU and Ubuntu 18.04 OS.
In particular, these VNFs of the CUPS-based CN, e.g., HSS, MME, SPGW-C, and SPGW-U, are implemented with Docker-based computing virtualization that enables dynamic instantiation and flexible resource provisioning.

\textbf{Transport domain manager.} 
We design the transport domain manager (TDM) to dynamically create, modify, and delete the transport slices by leveraging software-defined network (SDN) technology~\cite{hu2014survey}.
We design the TDM based on SDN controllers~\cite{medved2014opendaylight, zhu2019sdn} and use the \emph{meters} API in OpenFlow protocol~\cite{mckeown2008openflow} to manage the bandwidth for different slices.
The \emph{meters} API limits the maximum data rate of associated flows.
We develop the TDM based on OpenDayLight (ODL)~\cite{medved2014opendaylight} with OpenFlow 1.30.
We use a Ruckus ICX 7150-C12P as the SDN switch to connect the eNB/gNB and the CN, where each port of the switch has 1Gbps capacity.
For the sake of simplicity, the ODL controller and the TDM are implemented in the workstation computer.

\textbf{Edge domain manager.}
On developing the edge domain manager (EDM), we use Docker container technique~\cite{merkel2014docker} to virtualize the computing resources and provide isolation for edge servers.
The EDM can manage the resources of edge servers, e.g., CPU, RAM, Disk, and I/O, through Docker runtime configuration interfaces.
We deploy the EDM on the workstation computer and use \emph{docker update} command to update the CPU and RAM allocation.
The edge server of a slice is co-located in the slice's SPGW-U containers for the sake of simplicity.

\textbf{The OnSlicing agents.}
We develop OnSlicing agents with PyTorch 1.5, where all policy networks use 3-layer fully connected neural network with \emph{ReLU} activation functions, i.e., 128x64x32.
The activation functions of actor networks are \emph{Sigmoid}~\cite{goodfellow2016deep} to ensure that the action is between 0 and 1.
We deploy OnSlicing agents on the workstation computer to interact with slice tenants and the OnSlicing manager.

\section{Performance Evaluation}
\label{sec:evaluation}

\subsection{Evaluation Setups}

\textbf{Slice.} We develop three slices, and each hosts a mobile application with different resource demands and performance metrics.
We develop three mobile applications in Android and corresponded server applications using Python\footnote{The mobile applications are designed to utilize end-to-end resources and report diverse performance metrics periodically.}. 
The mobile applications can asynchronously send user requests, and the server applications serve user requests in parallel.
Thus, we can use one smartphone to emulate varying traffic with asynchronous user requests.

\textbf{MAR App.}
The MAR application continuously sends frames (540p) to the edge server and waits for the processing results.
The back-end server receives the frame, extracts the keypoint features with a feature extraction algorithm (ORB~\cite{rublee2011orb}), matches the features with a feature dataset, and returns the matched objects back to the phone.
The performance requirement of the MAR slice is the average round-trip latency of frames (500 ms). 
It can be recognized as a delay-sensitive application.

\textbf{HVS App.}
For HD video streaming (HVS), the stream server continuously streams 1080p video frames to mobile phones.
The performance requirement of the HVS slice is the average FPS of the streaming video (30 FPS).
It can be seen as a bandwidth-hungry application.

\textbf{RDC App.}
The reliable distant control (RDC) is developed to enable the remote control of wireless connected IoT devices.
The control server periodically receives raw data from users and sends the control message back to users.
We consider the size of both raw data and control message are 1 KBits for all users.
The phone is connected with a USB on-the-go (OTG) LED which indicates the message is received.
The performance requirement of the RDC slice is the reliability of radio transmission (99.999\%).
It can be identified as a reliability-sensitive application.

\textbf{Traffic Traces.}
We use an open mobile traffic dataset, i.e., Telecom Italia~\cite{barlacchi2015multi}, to generate the traffic of slices.
The data set consists of the Call, SMS, and Internet connections in thousands of base stations with minimum 10 minutes intervals over the Province of Trento, Italy.
We use the traffic trace of base stations as the slice traffic, where we scale the maximum traffic volume according to the capability of the testbed (5 users/s for MAR, 2 users/s for HVS, and 100 users/s for RDC).
With the arrival rate derived from traffic traces, we emulate the traffic of slices during the configuration interval (i.e., generating all arrival timestamp of users) according to the Poisson point process (PPP).
For example, if a MAR slice has average of 5 users/s traffic at the current time slot, the mobile application sends the frames to the server asynchronously, where the intervals of frames are sampled from an exponential distribution with a rate 5.
We compare the performance of different methods using two metrics, i.e., the resource usage (average used resources of all slices calculated in Eq.~\ref{eq:reward_calculation}) and the SLA violation (if the cumulative cost exceeds SLA threshold $C_{\max}=5\%$, i.e., lower than 95\% probability of SLA guarantee).

\textbf{Training.}
We use the testbed to emulate the resource orchestration with the network configuration interval of 15 minutes\footnote{Existing operational mobile networks usually take 15 minutes or more to collect network status and configure the network infrastructures.}.
A transition, episode, and epoch are defined as the state-action-reward-cost pairs, 96 transitions (24 hr), 1000 transitions, respectively.
In our experiments, with the subseconds action enforcement in domain managers and fast performance reporting in mobile applications, the OnSlicing agents can complete a transition in 5 seconds with the stable statistical performance of slices.
OnSlicing agents usually converge within less than 100 epochs (see Fig.~\ref{fig:result_safe_method}).

\textbf{Comparison Methods.}
We compare OnSlicing with the following methods:
1) \emph{Baseline}: we develop the baseline method to allocate end-to-end resources in three steps. 
First, each slice is offline evaluated within a small-scale testbed to identify key action factors, i.e., $[U_u, U_b, U_c]$, $[U_d, U_b]$ and $[U_m, U_s]$ are selected for the MAR, HVS, and RDC slice, respectively.
Second, a grid search with \emph{scikit-learn}~\cite{pedregosa2011scikit} is conducted to seek the minimum resource usage under different slice traffic to meet the slice's performance requirement.
Third, the over-requested resources in domain managers are scaled with the projection method.
2) \emph{Model\_Based}: we develop a model-based method by using approximated performance models in each slice.
The end-to-end latency and frame rate are formulated as $p_{MAR} = (f \cdot s) /U_u + l_s $~\cite{ran2018deepdecision} and $p_{HVS} =  U_u / (f \cdot s)$~\cite{liu2018dare}, respectively.
Here, $f, s, l_s, U_u$ are the slice traffic, frame bitrate, static latency and uplink radio bandwidth, respectively.
The reliability $p_{RDC}$ depends on the retransmission probability~\cite{jemmali2013bit} that varies with multiple factors, e.g., radio channel quality and MCS.
According to the measurement results in Fig.~\ref{fig:virtualization_an_reliability}, we determine the MCS offset $U_m=6, U_s=0$ to meet the RDC slice's performance requirement. 
The problem of minimizing the overall resource usage is solved by using the \emph{CVXPY} tool~\cite{agrawal2018rewriting}.
3) \emph{OnRL}: OnRL~\cite{zhang2020onrl} is an online DRL solution to optimize the video telephony, which allows DRL agents to learn from the real system and use a rule-based policy (refer to \emph{Baseline}) as a backup policy. We implement OnRL with modified reward, state, and action space for orchestrating cross-domain resources. We find that the native OnRL fails to meet the slices' SLA during the online learning phase as it merely allocates minimal resources to all slices. Thus, we adopt its fundamental idea with extra improvements. We supplement the reward sharping method to be aware of constraints and the projection method to deal with resource over-requesting situations in OnRL.

\begin{table}[!t]
    \begin{tabular}[b]{c |c c}\hline
       \textbf{Metric}     &  Avg. res. usage (\%)  &  Avg. SLA violation (\%)  \\ \hline
       \textbf{ OnSlicing}     & \textbf{20.19}  & \textbf{0.00} \\ 
       \textbf{ OnRL}     & 23.08  &  15.40  \\ 
       \textbf{ Baseline} & 52.18 & 0.00 \\ 
       \textbf{ Model\_Based} & 59.04 & 3.13 \\ \hline
    \end{tabular}
    \captionof{table}{ Test performances of methods}
\label{tb:overall_performance}
\end{table}

\subsection{Results Analysis}
\textbf{Overall performance.}
We show the test performance of different methods after the online learning phase completes in Table~\ref{tb:overall_performance}.
We observe that OnSlicing achieves the minimum average resource usage with 12.5\%, 61.3\%, and 65.8\% fewer usages as compared to OnRL, Baseline, and Model\_Based, respectively. 
This verifies the effectiveness of OnSlicing agents in handling real complex networks and minimizing resource usages.
Moreover, OnSlicing maintains zero SLA violation on average, which attributes to the novel methods of OnSlicing to guarantee slices' SLA during the online learning phase.
In contrast, OnRL has a higher resource usage than OnSlicing, and gets a worse performance on average SLA violation (15.40\%).
This can be attributed to its ineffectiveness in constraint awareness, poor distributed coordination among agents, and inefficiency of learn-from-scratch.
Furthermore, Model\_Based uses more resources than Baseline and shows a larger SLA violation, which can be attributed to the inaccurate mathematical models that cannot fully represent the complex end-to-end network.

\begin{figure}[!t]
	\centering
	\includegraphics[width=3.1in]{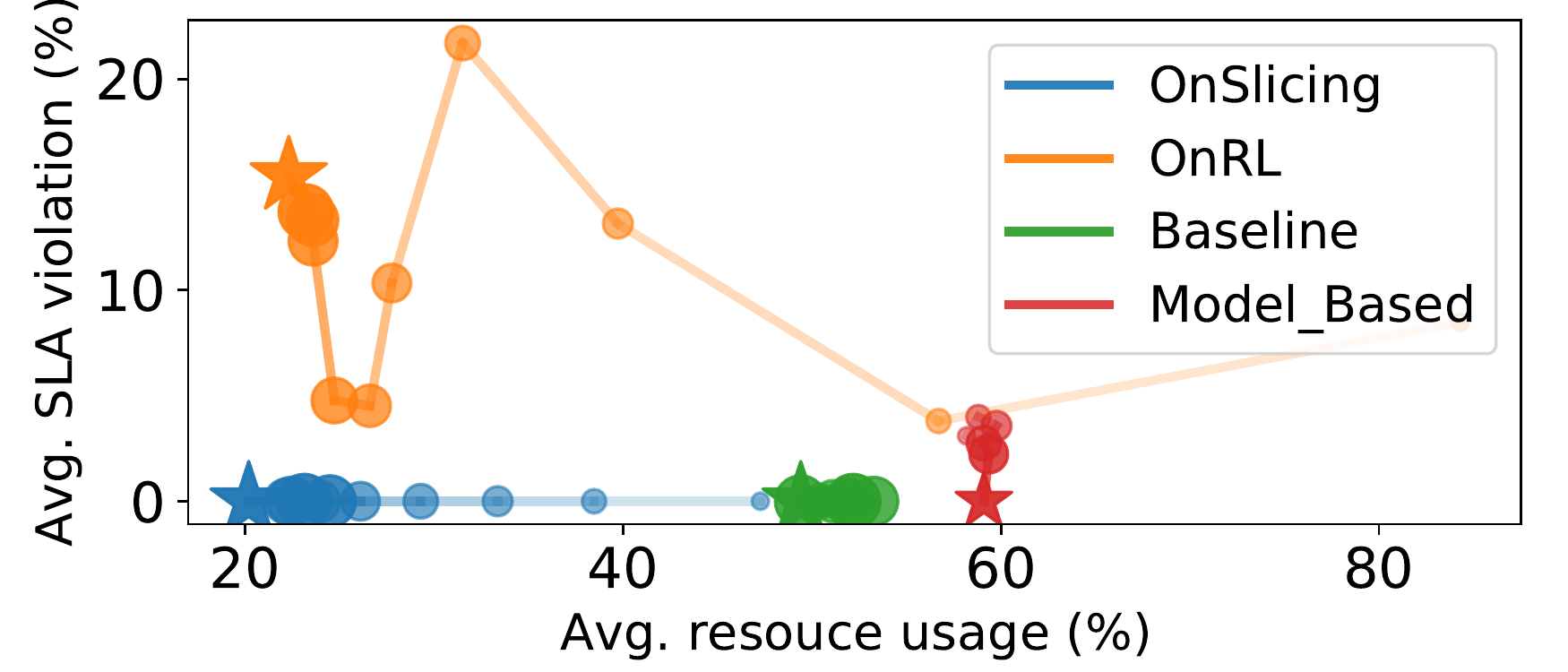}
	\caption{\small Learning trajectory of methods}
	\label{fig:result_algs_learning}
\end{figure}

\textbf{Online learning performance.}
Fig.~\ref{fig:result_algs_learning} shows the learning trajectory of different methods throughout the online learning phase.
Here, we denote the start and end point of a method with a small point marker and a large star marker, respectively.
We observe that OnRL starts with very high resource usage and SLA violation because the DRL agent needs to learn from scratch, and its average SLA violations change dramatically during the online learning phase.
In contrast, the average resource usage of OnSlicing gradually decreases without noticeable violations of slices' SLA.

\begin{figure*}[!t] 
\captionsetup{justification=centering}
  \begin{minipage}[t]{0.33\textwidth}
    \centering
    \includegraphics[width=2.3in, height=1.5in]{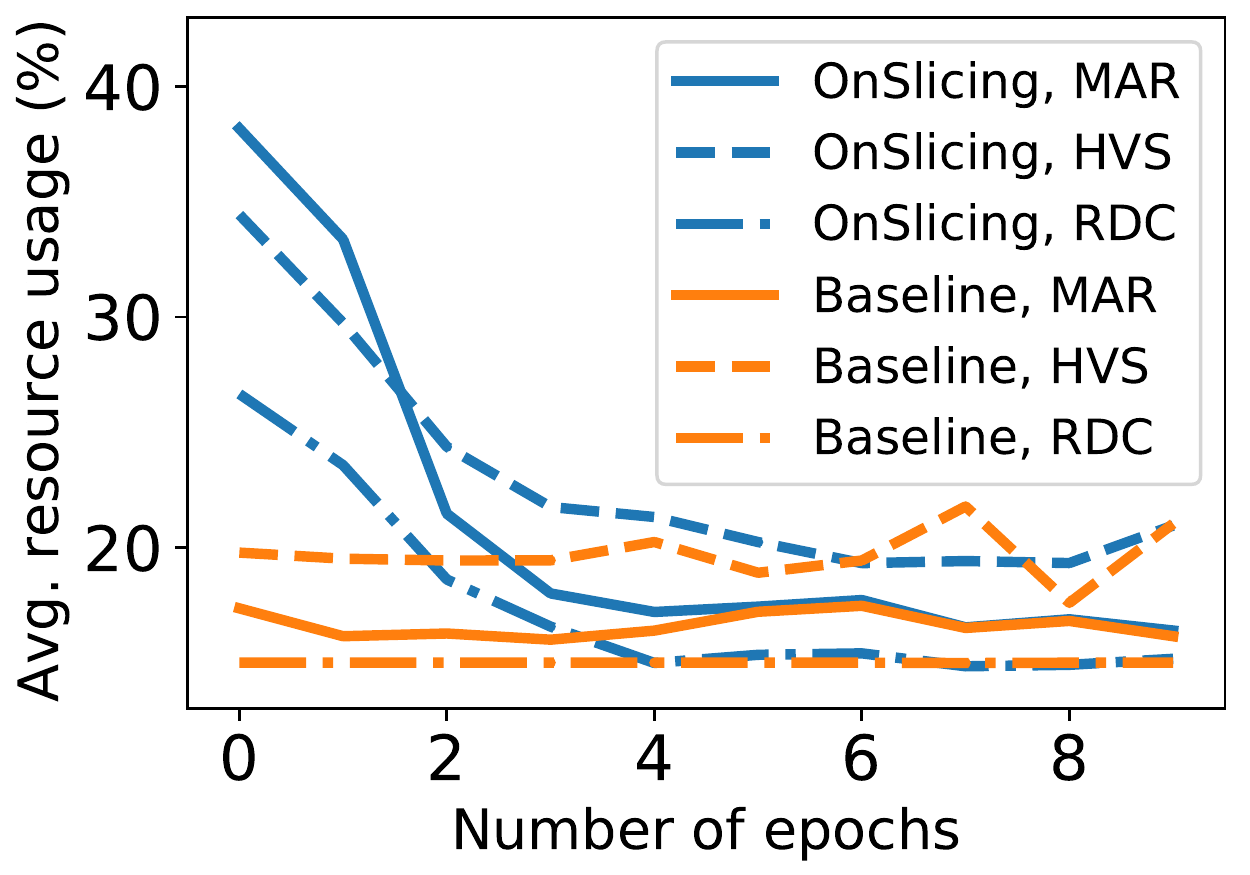}
    \captionof{figure}{\small Offline imitate learning from baseline}
    \label{fig:result_imitate_learning}
  \end{minipage}
  \hfill
  \begin{minipage}[t]{0.33\textwidth}
    \centering
    \includegraphics[width=2.3in, height=1.5in]{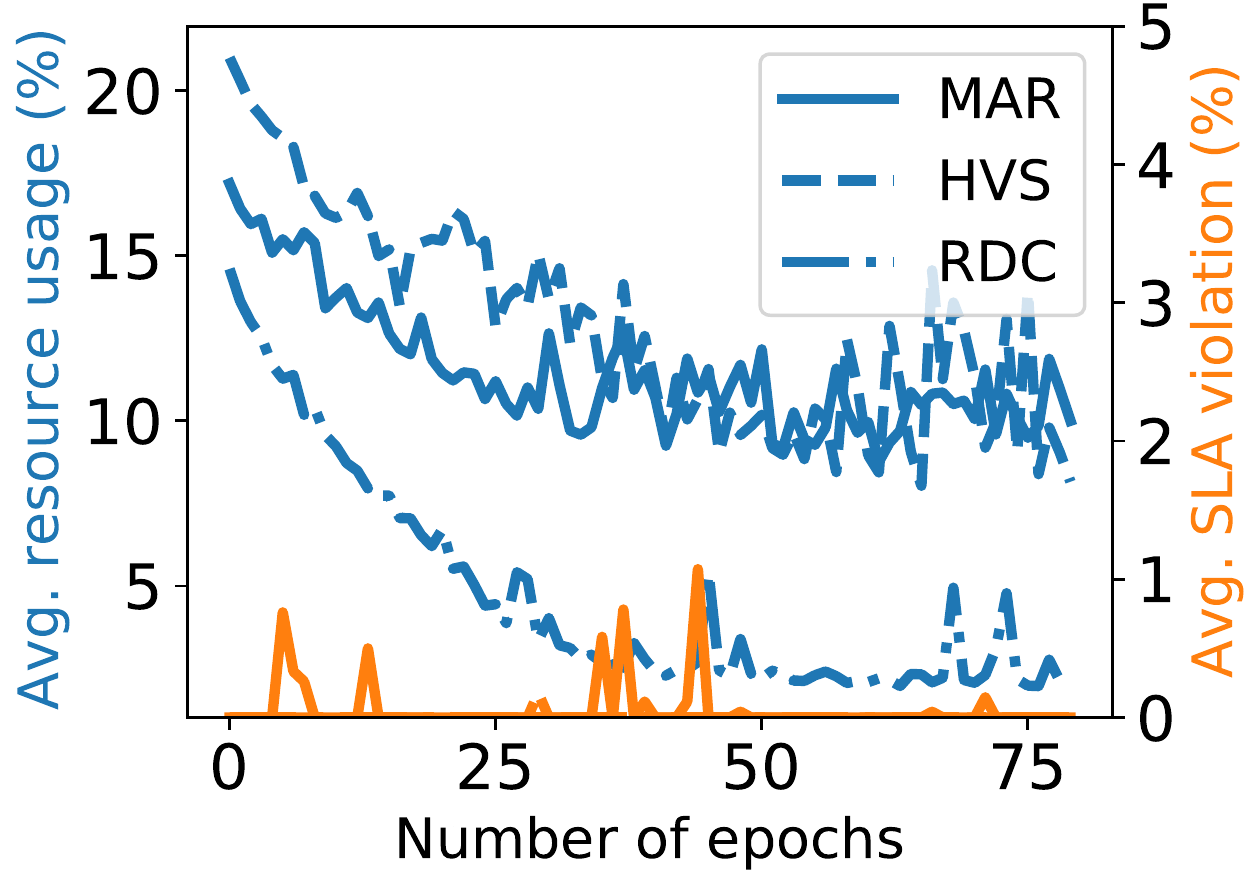}
    \captionof{figure}{\small Online learning of OnSlicing agents}
    \label{fig:result_onslicing_learning}
  \end{minipage}
  \hfill
  \begin{minipage}[t]{0.33\textwidth}
    \centering
    \includegraphics[width=2.3in, height=1.5in]{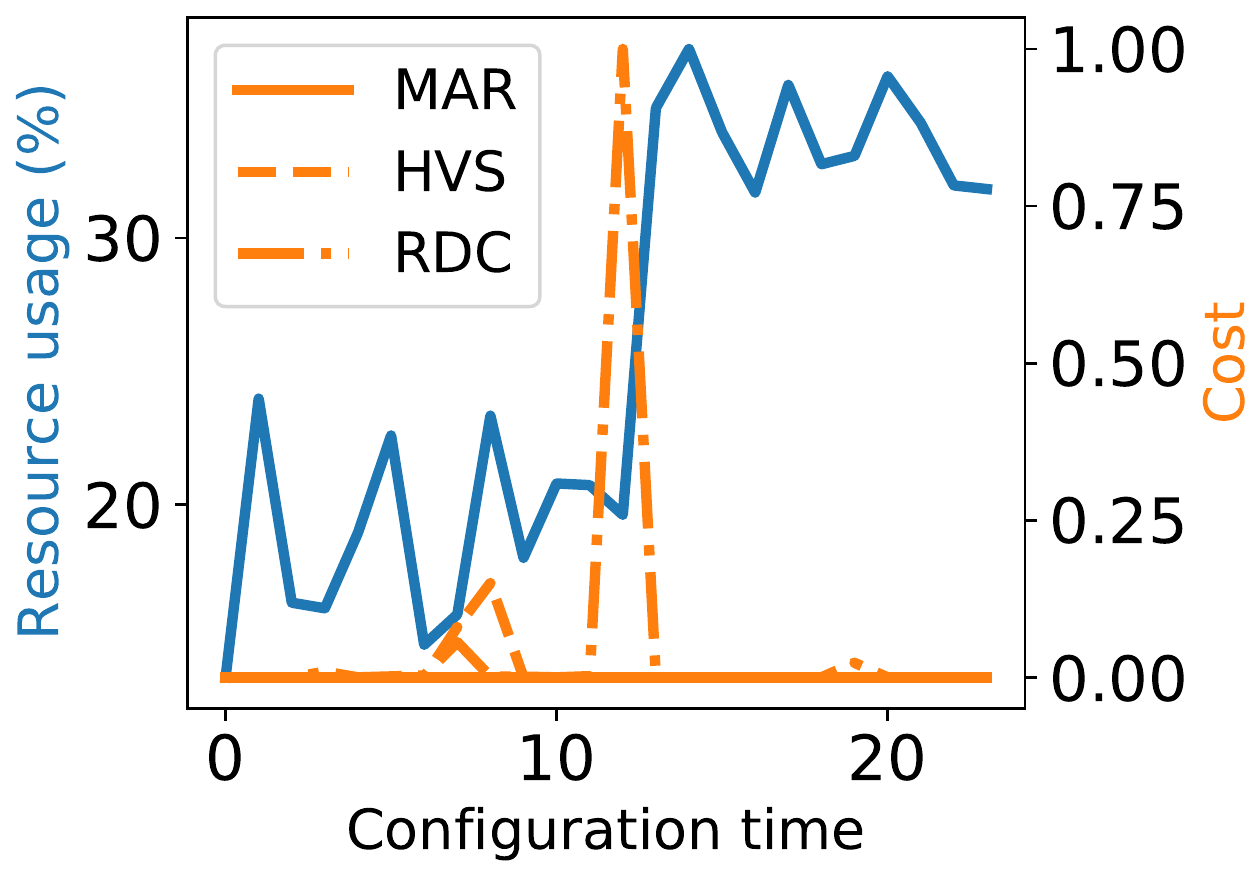}
    \captionof{figure}{\small A showcase of proactive baseline switching}
    \label{fig:result_safe_showcase}
  \end{minipage}
  \hfill
\end{figure*}

The remarkable resource usage reduction of OnSlicing can be attributed to, \emph{(i)} the offline imitate learning scheme from Baseline, and \emph{(ii)} the individualized learning for each slice. 
Fig.~\ref{fig:result_imitate_learning} shows the offline training curve of OnSlicing agents, where the average resource usage obtained by the agents gradually approach that of Baseline as they imitate the resource orchestration behaviors of Baseline.
Thus, OnSlicing agents can start online learning with a policy approximating Baseline.
Meanwhile, the OnSlicing agent in each individual slice could efficiently learn the unique characteristics of slice application, because of the low complexity of the individualized problem as compared to the problem of joint orchestration for all slices.

The low SLA violation of OnSlicing comes from, \emph{(i)} the constraint-aware policy update method, and \emph{(ii)} the proactive baseline switching mechanism.
Fig.~\ref{fig:result_onslicing_learning} shows the online learning curve of OnSlicing agents, where the average resource usage decreases gradually with near-zero SLA violations.
As OnSlicing incorporates the violation of slices' SLA into the reward function, the high-cost orchestration actions can be avoided.
As a result, OnSlicing achieves only several spikes of violation (maximum 1\%) throughout the online learning phase.
Besides, OnSlicing agents can switch to Baseline proactively, which prevents SLA violations caused by the action exploration.
We illustrate the proactive baseline switching mechanism in Fig.~\ref{fig:result_safe_showcase}, where Baseline is triggered as an abnormal spike of violation happens in the HVS slice (at time slot 12), and thus the resource usage increases from $\sim$20\% to $\sim$35\% consequently. 
The proactive baseline switching mechanism relies on policy $\pi_\phi$ to predict the cost value function of Baseline, where the predictions may be inaccurate due to unseen states.
In this situation, Baseline is invoked late and the performance requirement of a slice could be violated slightly as shown in Fig.~\ref{fig:result_safe_showcase}.

\begin{figure*}[!t] 
\captionsetup{justification=centering}
  \begin{minipage}[t]{0.33\textwidth}
    \centering
    \includegraphics[width=2.3in, height=1.5in]{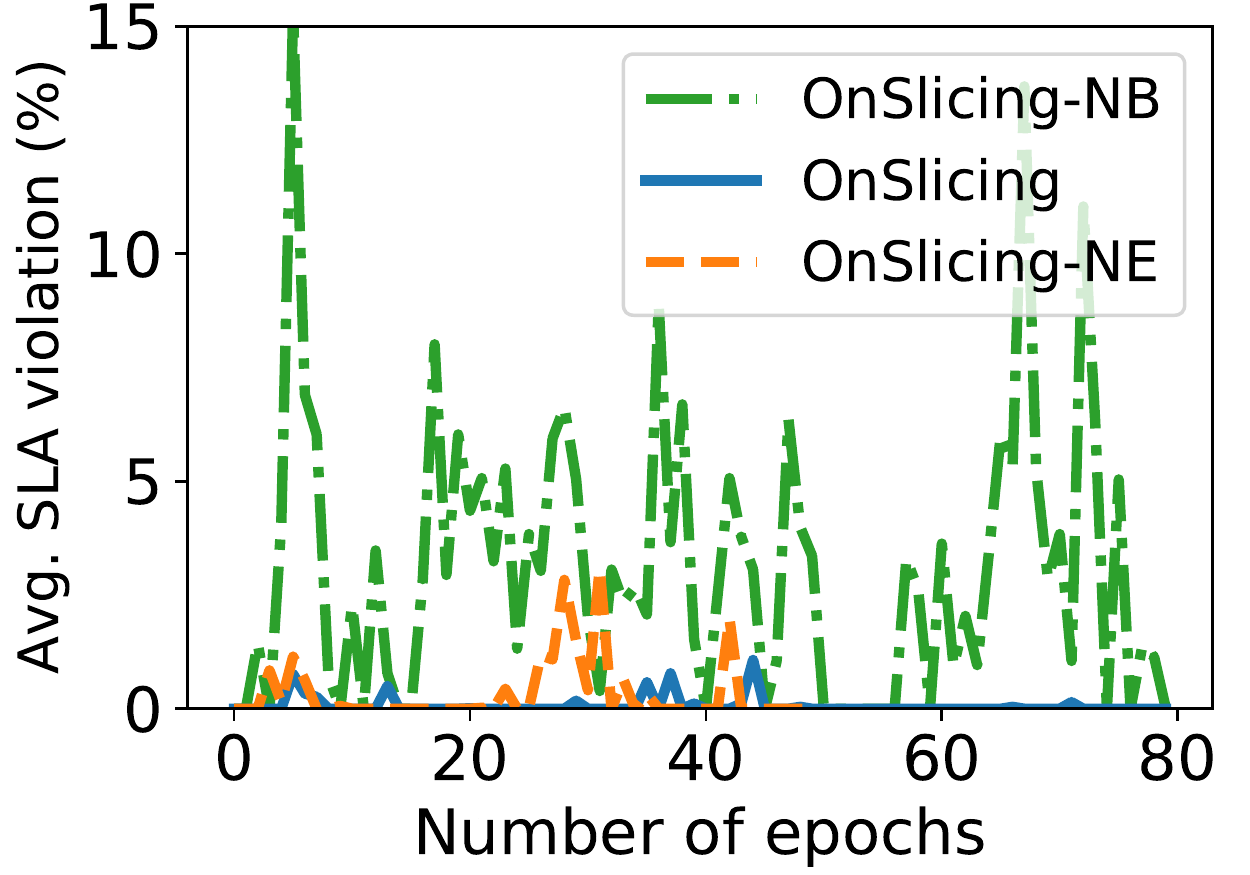}
    \caption{\small Performance of baseline switching methods}
    \label{fig:result_safe_method}
  \end{minipage}
  \begin{minipage}[t]{0.33\textwidth}
    \centering
    \includegraphics[width=2.3in, height=1.5in]{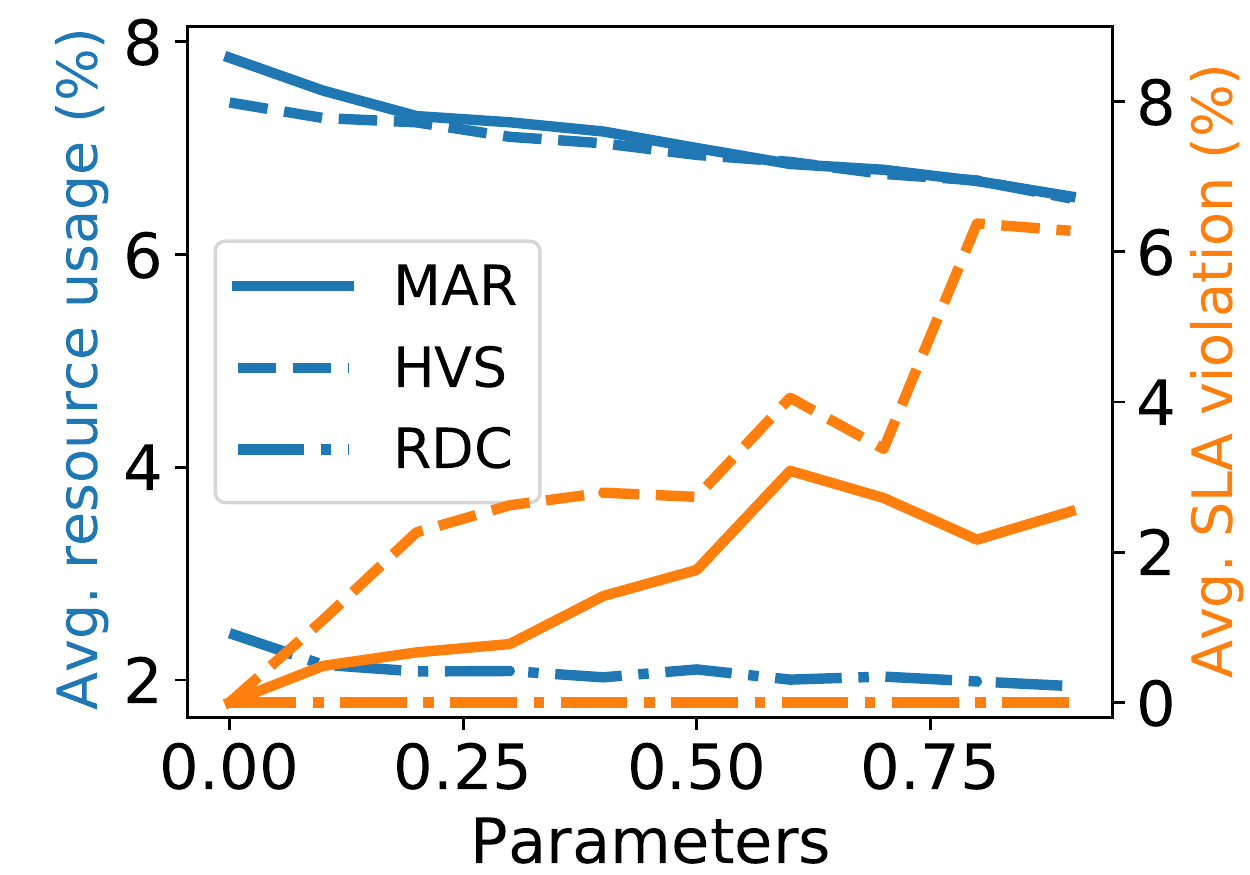}
    \captionof{figure}{\small Resource usage under coordinating parameters}
    \label{fig:result_scalable_dual}
  \end{minipage}
  \hfill
  \begin{minipage}[t]{0.33\textwidth}
    \centering
    \includegraphics[width=2.3in, height=1.5in]{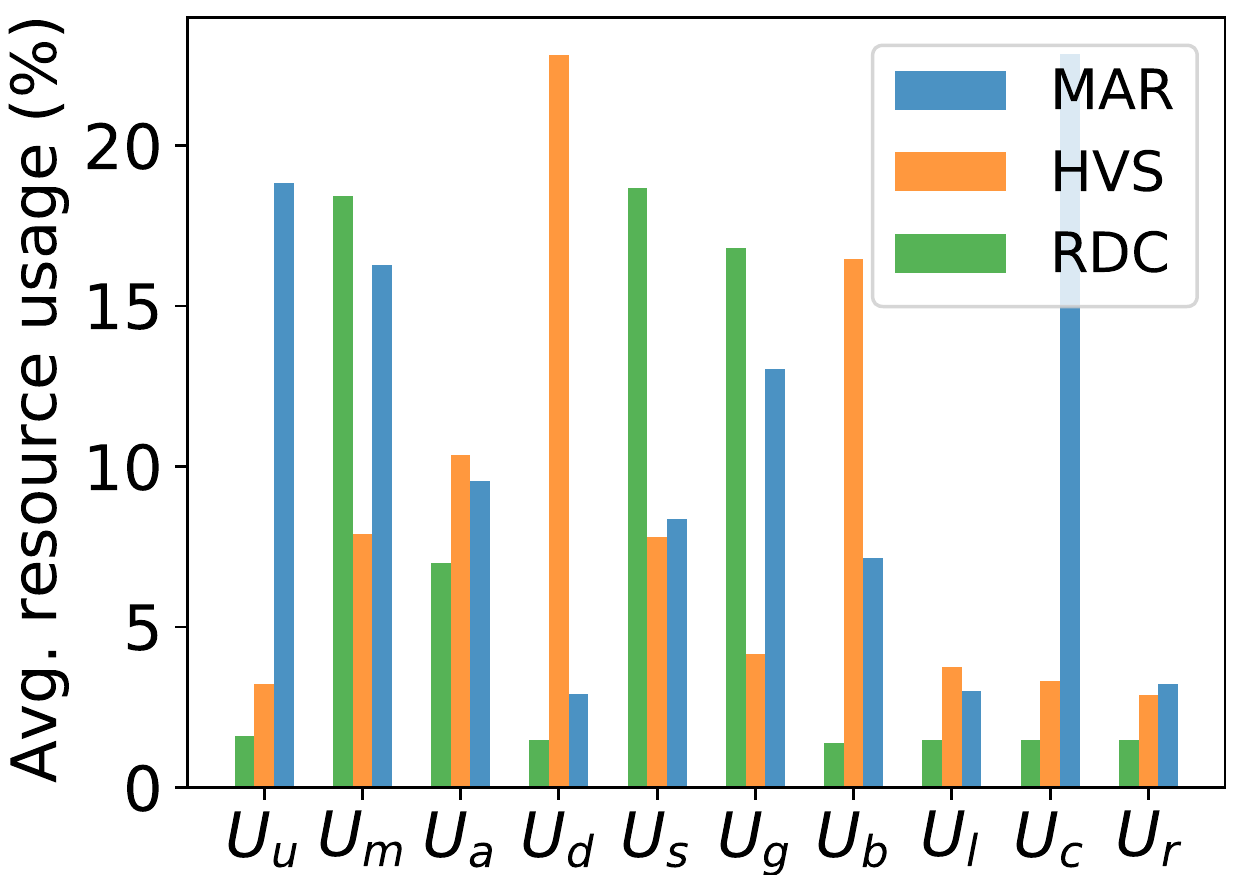}
    \captionof{figure}{\small The avg. used resource of slices}
    \label{fig:result_imitate_action}
  \end{minipage}
\end{figure*}

\begin{table}[!t]
\small
    \begin{tabular}[b]{c|c c c}\hline
       \textbf{Method}     &  Avg. res. usage (\%)  &  Avg. SLA viol. (\%) \\ \hline
       \textbf{ OnSlicing}     & 29.07  & 0.06 \\ 
       \textbf{ OnSlicing-NE}     &  30.81  &  0.33  \\ 
       \textbf{ OnSlicing-NB} & 29.64 & 2.94 \\ 
       \textbf{ OnSlicing Est. Noise} & 52.91 & 1.03 \\ \hline
    \end{tabular}
    \captionof{table}{\small Avg. performance of baseline switching methods}
\label{tb:baseline_switching}
\end{table}

\textbf{Learning with near-zero violations.}
Fig.~\ref{fig:result_safe_method} shows the average SLA violation of OnSlicing with different baseline switching mechanisms, i.e., OnSlicing-NB (non-baseline) and OnSlicing-NE (non-estimator), during the online learning phase.
Here, OnSlicing-NB does not switch to Baseline while OnSlicing-NE switches to Baseline only if the cumulative costs exceed the given SLA threshold $C_{\max}$ without predicting the cost value function.
The experimental results show that OnSlicing-NB has the worst performance in terms of the average SLA violation (2.94\%) because no Baseline can be switched to when the OnSlicing-NB agent violates the performance requirement of slices.
OnSlicing-NE obtains a less SLA violation (0.33\%) but a higher resource usage than OnSlicing-NB because it switches to Baseline that has higher resource usages.
Table~\ref{tb:baseline_switching} shows the average resource usage and SLA violation achieved by these mechanisms throughout the online learning phase.
Although OnSlicing-NB and OnSlicing-NE achieve similar average resource usage, their high average SLA violations suggest the necessity of the proactive baseline switching mechanism in OnSlicing.
Besides, we manually add a large Gaussian noise with 1.0 variance on the output of policy $\pi_\phi$ to emulate its prediction error of the cost value function and evaluate the robustness of OnSlicing. The average resource usage of OnSlicing Est. Noise is substantially worse than that of OnSlicing but is similar as that of Baseline. This is because the baseline switching mechanism reacts when the cumulative cost violates the SLA threshold (Eq.~\ref{eq:switching_criteria}), even if the estimator generates inaccurate predictions. 

\begin{table}[!t]
\small
    \begin{tabular}[b]{c |c c c}\hline
       \textbf{Methods}              &  Usage (\%)    &  Viol. (\%)  &  Interact num.  \\ \hline
       \textbf{OnSlicing}     & 20.2 $\pm$ 0.23      & 0.00 $\pm$ 0.00  & 1.83 $\pm$ 0.61 \\ 
       \textbf{OnSlicing-projection}      & 18.2 $\pm$ 0.50      & 3.66 $\pm$ 2.49  & 1.00 $\pm$ 0.00 \\  
       \textbf{OnSlicing Md. Noise}      & 23.8 $\pm$ 1.56      & 2.57 $\pm$ 1.66  & 2.16 $\pm$ 1.08 \\  \hline
    \end{tabular}
    \captionof{table}{\small Performance of action modifications}
\label{tb:scalable_performance}
\end{table}




%
\textbf{Learning in distributed networks.}
We compare the performance of OnSlicing under different action modification methods in Table~\ref{tb:scalable_performance}.
Although the OnSlicing-projection method obtains a slightly lower resource usage than OnSlicing, it incurs a much higher SLA violation because the over-request resources are scaled down which results in an under-provisioned resource of slices.
OnSlicing only needs 1.83 times interactions between OnSlicing agents and domain managers on average, which verifies the effectiveness of the parameter initialization in consecutive time slots.
Besides, we manually add a large Gaussian noise with 1.0 variance on the output of the action modifier to emulate its failure of action modification. Although OnSlicing Md. Noise has an increment in both the resource usage and SLA violation, its SLA violation is still lower than that of OnSlicing-projection method, which verifies the robustness of OnSlicing.
Meanwhile, we evaluate the performance of slices under fixed coordinating parameters $\beta^k_t, \forall k \in \mathcal{K}$, on all resources in Fig.~\ref{fig:result_scalable_dual}.
We find that the average resource usage decreases as the increase of parameters on all resources, which validates that the action modifier in OnSlicing can adjust the resource orchestration according to the guide of domain managers.

Besides, we show that the average orchestrated resources generated by OnSlicing agents for different slices in Fig.~\ref{fig:result_imitate_action}.
It can be seen that OnSlicing agents learn the inherent characteristics of different slices in terms of resource demands through the online learning phase.
For example, the MAR slice is allocated more uplink radio resources $U_u$ and computing resources $U_c$,
the HVS slice is allocated more downlink radio resources $U_d$, and the RDC slice is allocated higher MCS offsets $U_m, U_s$ for both uplink and downlink.

\begin{figure}[!t] 
\captionsetup{justification=centering}
  \begin{minipage}[t]{0.235\textwidth}
    \centering
    \includegraphics[width=1.65in, height=1.3in]{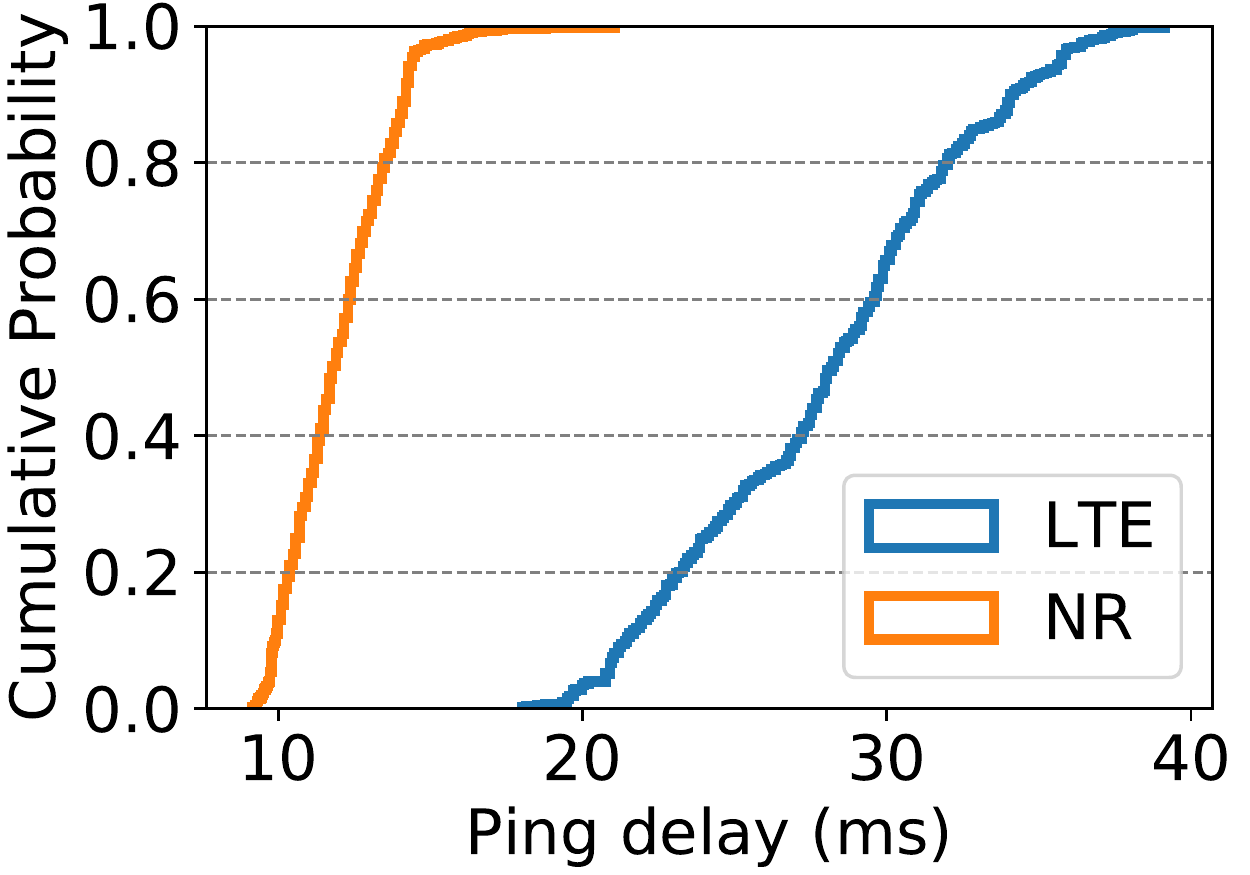}
    \captionof{figure}{\small Ping delay in LTE and NR}
    \label{fig:result_5g_delay}
  \end{minipage}
  \hfill
  \begin{minipage}[t]{0.235\textwidth}
    \centering
    \includegraphics[width=1.65in, height=1.3in]{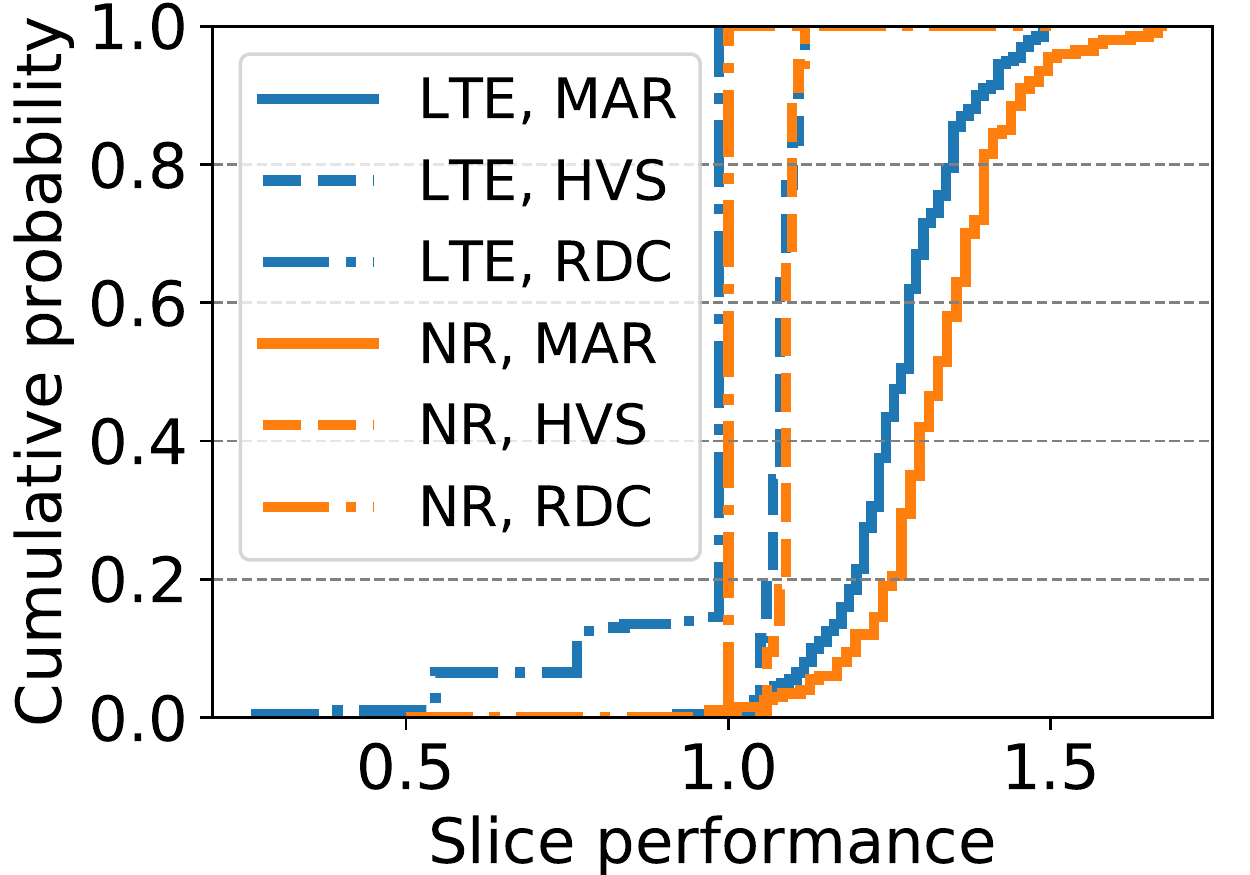}
    \captionof{figure}{\small Individual slice performance in LTE and NR}
    \label{fig:result_5g_slice_performance}
  \end{minipage}
\end{figure}

\textbf{Performance in 5G.}
We evaluate the performance of OnSlicing in a 5G NSA scenario, in which gNB uses 40MHz bandwidth with total 106 PRBs (30kHz subcarrier spacing).
The TDD configuration is 5 slots and 6 symbols for downlink, and 4 slots and 4 symbols for uplink.
For stabilizing the 5G experiments, we set a fixed MCS 9 for both uplink and downlink. We apply the fixed MCS for 4G LTE experiments for a fair comparison.
Under this scenario, the average throughput measured by \emph{iperf3} are 6.71 Mbps UL and 14.3 Mbps DL in 4G LTE, and 11.5 Mbps UL and 18.5 Mbps DL in 5G NR.

We compare the ping delay between smartphones and SPWG-Us as shown in Fig.~\ref{fig:result_5g_delay}.
We observe that 5G NR (avg. 11.99 ms) achieves a substantial reduction of ping delay than 4G LTE (avg. 27.99 ms), which is attributed to RAN improvements as we use the identical TN and CN for both 4G LTE and 5G NR.
The significant delay reduction as well as the higher data rate of 5G NR help to improve the performance of various applications.
We show CDF of the slices' performance, i.e., $p_t(\mathbf{s}_t, \mathbf{a}_t)/P$, in Fig.~\ref{fig:result_5g_slice_performance}.
We observe that 5G NR achieves noticeable performance improvement on both MAR slice (avg. latency) and RDC slice (reliability).
Meanwhile, the performance of the HVS slice under 4G LTE and 5G NR are similar because the streaming server streams to users with a fixed frame rate and does not saturate the downlink bandwidth.
Furthermore, we show the performance of OnSlicing in both 4G LTE and 5G NR in Table~\ref{tb:5g_performance}.
Considering the fixed MCS setting in the experiment, more radio resources are needed to meet the requirements of slices, and thus the average resource usage for both 4G LTE and 5G NR are increased accordingly.
Meanwhile, there is a slight average SLA violation in 4G LTE since the limited uplink and downlink bandwidth cannot handle peak traffic of slices.
In contrast, OnSlicing in 5G NR achieves zero violation, which attributes to the high data rate and low delay in RAN.

\begin{figure}[!t] 
\captionsetup{justification=centering}
  \begin{minipage}[t]{0.235\textwidth}
    \centering
    \includegraphics[width=1.6in, height=1.3in]{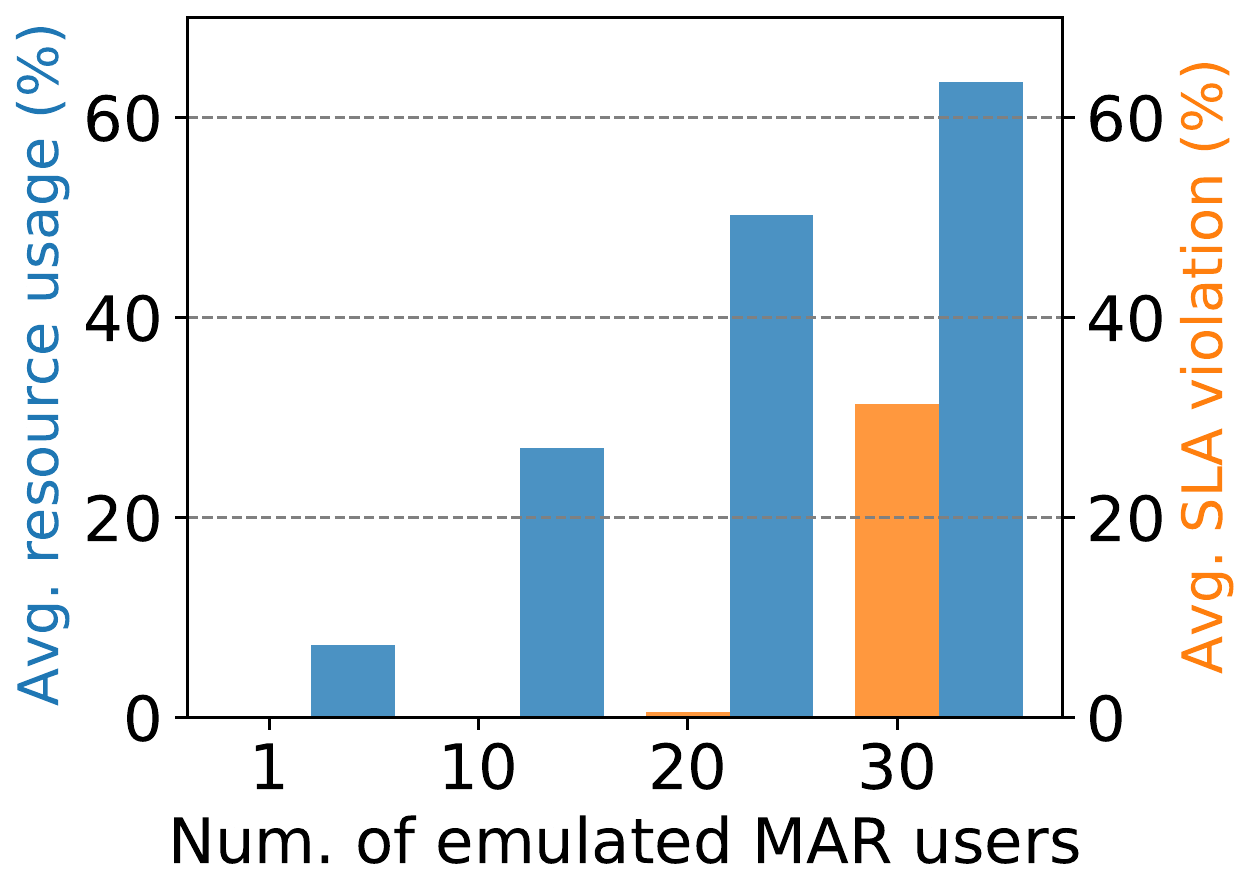}
    \captionof{figure}{\small Performance under varying user numbers}
    \label{fig:result_emulation_users}
  \end{minipage}
  \hfill
  \begin{minipage}[t]{0.235\textwidth}
    \centering
    \includegraphics[width=1.6in, height=1.3in]{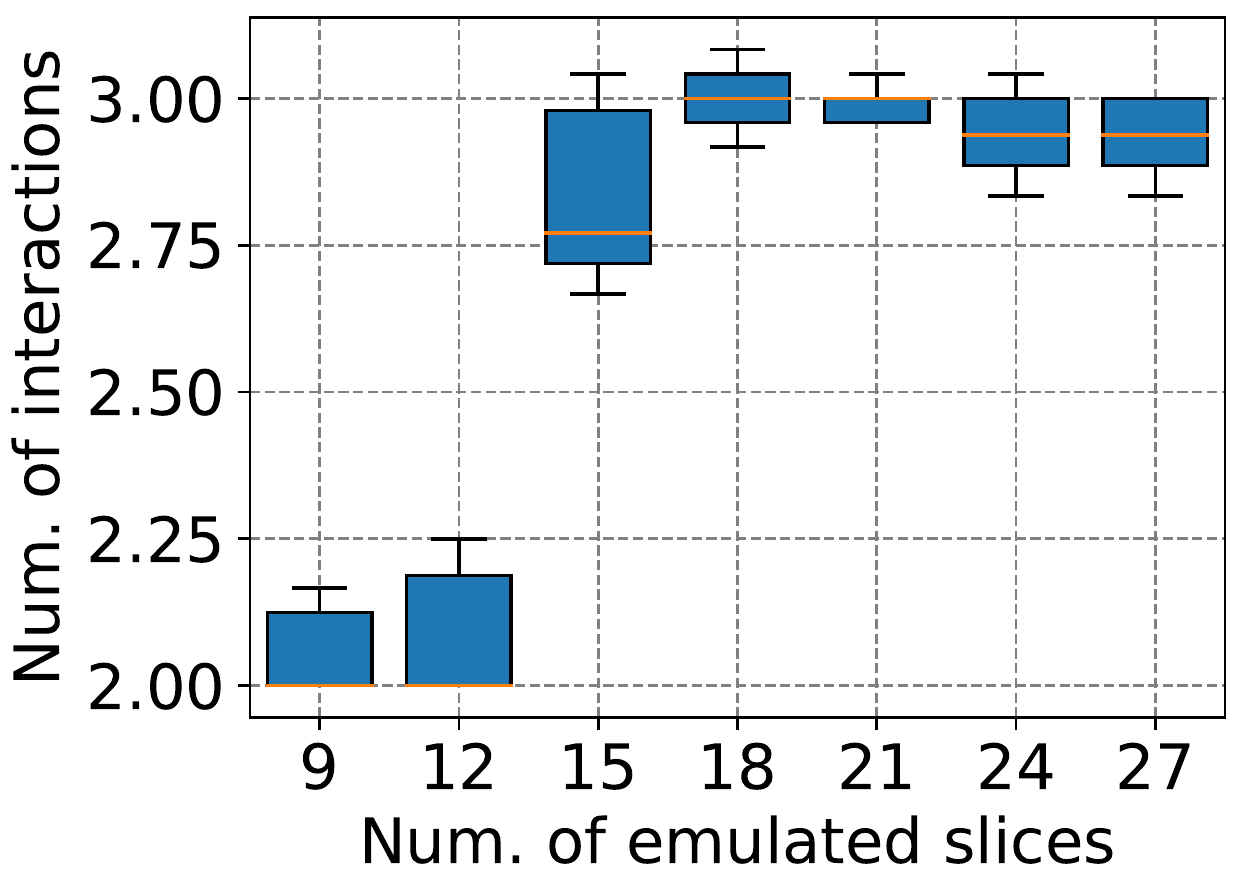}
    \captionof{figure}{\small Performance under varying slice numbers}
    \label{fig:result_emulation_slices}
  \end{minipage}
\end{figure}

\textbf{Learning in large-scale.}
We evaluate the performance of OnSlicing in large-scale emulation, in which mobile users are emulated with the OAI platform to connect LTE eNB with the L2 network-FAPI (nFAPI) interface. In particular, the applications in emulated users send the traffic to edge servers through the emulated Ethernet ports (e.g., \emph{oai-ue1}). The rest of the testbed, e.g., RAN, TN, CN, and EN, are the same as compared to previous experiments. Although the transmission and processing below layer 2 are omitted, the emulation platform can emulate radio channel dynamics by varying the capacity per PRB in carrying user data\footnote{We keep the radio channel quality constant in the emulation as 1) we evaluate the large-scale performance regarding multi-agent interaction, 2) we keep the stability of the emulation platform.}.
As shown in Fig.~\ref{fig:result_emulation_users}, the average resource usage of OnSlicing is increased when there are more users in the MAR slice.
Meanwhile, the average SLA violation is maintained low until the system is overwhelmed by a massive number of slice users.
It is worth to mention that the slice agent does not need to be retrained when dealing with varying slice traffic.
In addition, we show the average number of interactions between OnSlicing agents and domain managers as the number of slices increases in Fig.~\ref{fig:result_emulation_slices}.
It can be seen that the number of interactions is kept low, e.g., 3 times, which verifies that OnSlicing can scale to orchestrate cross-domain resources in large-scale networks.


\begin{table}[!t]
    \begin{tabular}[b]{c |c c }\hline
       \textbf{Networks}              &  Avg. res. usage (\%)    &  Avg. SLA violation (\%)  \\ \hline
       \textbf{5G NR}     & 43.5 $\pm$ 3.27      & 0.00 $\pm$ 0.00   \\ 
       \textbf{4G LTE}    & 45.9 $\pm$ 4.48      & 0.66 $\pm$ 1.42   \\ \hline
    \end{tabular}
    \captionof{table}{ OnSlicing performance in 4G LTE and 5G NSA}
\label{tb:5g_performance}
\end{table}

\section{Related Work}

\textbf{Network slicing management:}
Network slicing is the key technique to cost-efficiently support heterogeneous use cases and services~\cite{foukas2016flexran, bega2019machine, salvat2018overbooking, d2020sl}.
Orion as the first RAN slicing solution~\cite{foukas2017orion} enables dynamic on-the-fly virtualization of base stations, which is developed based on the FlexRAN platform~\cite{foukas2016flexran}.
Marqueze \emph{et. al.}~\cite{marquez2018should} showed the empirical study of resource management efficiency in network slicing, which advocates the dynamic orchestration of cross-domain resources.
Salvat \emph{et. al.}~\cite{salvat2018overbooking} proposed two resource provisioning algorithms that maximize the revenue of MNO in network slicing, where slices are identified by different PLMN-Ids which is not dynamic.
Multiple slice admission and resource provisioning algorithms~\cite{han2019utility, d2020sl} are proposed to further improve the efficiency and performance of network slicing systems. 
However, these works cannot provide dynamic end-to-end slicing including RAN, TN, CN, and EN.
Also, they formulate problems with approximated mathematical models, which suffer from the discrepancy between these models and real networks.
In contrast, OnSlicing is a model-free approach and enables online learning to automatically adapt to real networks.

\textbf{Machine learning for networking:}
Machine learning techniques have been increasingly studied to deal with complex and time-correlated network systems~\cite{lee2020perceive, yan2020learning, wang2020job, aggarwal2020libra}.
EdgeSlice~\cite{liu2020edgeslice} uses a decentralized DRL approach to orchestrate cross-domain resources in distributed networks to meet slices' SLA.
Bega \emph{et. al.}~\cite{bega2019deepcog} proposed DeepCog with deep learning techniques to predict network capacity within individual slices and balance the tradeoff between resource over-provisioning and service request violations.
Microscope~\cite{zhang2020microscope} enables efficient service demand estimation of slices by proposing a deep learning-based decomposition technique to deal with complex spatiotemporal features hidden in traffic aggregates.
Several works~\cite{khairy2020constrained, liu2020constrained,liu2020edgeslice} used constrained DRL approaches to satisfy the constraints in network management by using the reward shaping and Lagrangian primal-dual methods.
To bridge the simulation-to-reality gap, OnRL~\cite{zhang2020onrl} allows online DRL within real networks, which improves the performance of the real-time mobile video telephony by proposing an individualized hybrid learning algorithm and a learning aggregation mechanism.
However, these works rely on learning algorithms with unconstrained exploration, which cannot comply with system constraints in infrastructures and could violate slices' SLA during the online learning phase.
OnSlicing introduces the constraint-aware policy update method, proactive baseline switching mechanism and distributed coordination among agents, which achieves near-zero violations of slices' SLA and maintains distributed system limitations.

\section{Discussion}

\textbf{Scalability.}
OnSlicing is implemented and evaluated using a small-scale system testbed, and its performances are verified using extensive experiments.
We design OnSlicing with the consideration of practical large-scale deployment in operational networks.
For example, we create an individualized agent for each slice, which can seamlessly scale to support hundreds or thousands of slices in future. In contrast, the centralization of resource orchestration with a single agent fails to scale and adapt to network topology changes.
In addition, we develop the distributed coordination mechanism to maintain resource constraints in infrastructures, which can be applied to the different number of agents and domain managers.
The overheads incurred by OnSlicing is small, in terms of virtualization of infrastructures, state collection and action enforcement, and the execution of OnSlicing agents.
As OnSlicing is deployed in a large-scale network, a potential challenge may arise when the state space is extended to be extremely large, and the action space turns significantly heterogeneous in terms of enforcement delay.
As a result, the collection of states may aggregate the traffic burden of transport networks and the enforcement of actions could lead to imbalance delay by domain managers (e.g., PRB allocation in RAN needs millisecs while server scaling in EN requires seconds).

\textbf{Exploration.}
We design OnSlicing to stop exploration, i.e., baseline switching, when we predict its failure in slices' SLA assurance.
Although it might slow down the online learning progress of OnSlicing agents toward the optimal policy, OnSlicing becomes safer in the meantime (e.g., near-zero SLA violations).
The exploration of OnSlicing is mainly controlled by the factor $\eta$ in Eq.~\ref{eq:switching_criteria} and the SLA threshold $C_{\max}$.
With the larger $\eta$ and the smaller SLA threshold, OnSlicing agents are more conservative and switch to baseline earlier.

\textbf{Dynamics.}
OnSlicing agents are trained under various network dynamics, which allows them to make appropriate orchestration actions under different states.
For example, although the positioning of smartphones and base stations are stationary, moderate variations of radio channel conditions of slice users are observed in experiments.
Besides, we emulate traffic variations in slices during the online learning phase, where each slice may have different traffic patterns and volumes. 
As OnSlicing is deployed in operational networks, we may see more dynamics such as new traffic pattern of slices, and expect OnSlicing agents to adapt to new dynamics via online learning.

\textbf{Convergence.}
In general, DRL agents require a large number of transitions to learn the optimal policy.
This can arise an issue because the network orchestration in existing operational networks normally happens at the timescale of hours rather than seconds.
As a result, online learning could take weeks or even months to achieve the optimal policy.
In OnSlicing, the agents offline imitate the rule-based policy, and then keep learning and improving the policy performance smoothly during the online learning phase.
In other words, OnSlicing always performs better than the rule-based policy, which helps mitigate this issue.
Furthermore, several promising techniques could accelerate the learning progress, e.g., policy aggregation~\cite{zhang2020onrl} and federated learning~\cite{bonawitz2019towards}, which can be further incorporated into OnSlicing.



\section{Conclusion}
In this work, we have designed OnSlicing, an online end-to-end network slicing system.
We addressed multiple practical challenges of online DRL-based resource orchestration including the performance assurance of slices and scalability in distributed networks.
The experimental results validated that OnSlicing achieves the minimum cross-domain resource usage with near-zero violations of slices' SLA throughout the online learning phase. 
OnSlicing shed the light on incorporating online DRL into network management in next-generation mobile networks.

\begin{acks}

This work is partially completed at Nokia Bell Labs. Dr. Tao Han's work is partially supported by the US National Science Foundation under Grant No. 2147821, No. 2147623, No. 2047655, and No. 2049875.

We would like to thank the shepherd, Prof. Sangeetha Abdu Jyothi for the guidance in refining this paper and anonymous reviewers for their insightful comments.
\end{acks}

\bibliographystyle{acm}
\bibliography{ref/reference}

\end{document}